\def\year{2021}\relax
\documentclass[letterpaper]{article} 
\usepackage{aaai21}  
\usepackage{times}  
\usepackage{helvet} 
\usepackage{courier}  
\usepackage[hyphens]{url}  
\usepackage{graphicx} 
\usepackage{csquotes}
\usepackage{rotating}
\usepackage{natbib}  
\usepackage{caption} 
\usepackage{array}
\usepackage{setspace} 
\usepackage{makecell}
\usepackage[flushleft]{threeparttable}
\usepackage{booktabs,caption}
\usepackage[toc,page]{appendix}
\usepackage{subcaption}
\usepackage{rotating}
\usepackage{multirow}

\usepackage{amssymb}

\usepackage{makecell}
\usepackage{url}
\usepackage{xurl}

\title{Social Media Study of Public Opinions on Potential COVID-19 Vaccines: Informing Dissent, Disparities, and Dissemination}
\author{
    Hanjia Lyu, \textsuperscript{\rm 1}
    Junda Wang, \textsuperscript{\rm 2}
    Wei Wu, \textsuperscript{\rm 1}
    Viet Duong, \textsuperscript{\rm 2}\\
    Xiyang Zhang, \textsuperscript{\rm 3}
    Timothy D. Dye, \textsuperscript{\rm 4}
    Jiebo Luo \textsuperscript{\rm 2,*}
    \\
}
\affiliations{
    \textsuperscript{\rm 1} University of Rochester, Goergen Institute for Data Science, Rochester\\
    \textsuperscript{\rm 2} University of Rochester, Department of Computer Science, Rochester\\
    \textsuperscript{\rm 3} University of Akron, Department of Psychology, Akron\\
    \textsuperscript{\rm 4} University of Rochester School of Medicine and Dentistry, Department of Obstetrics and Gynecology, Rochester\\

    \textsuperscript{\rm *} Corresponding author: Jiebo Luo (jluo@cs.rochester.edu)
}

\begin{document}

\maketitle

\begin{abstract}
\textbf{Background:} The current development of vaccines for SARS-CoV-2 is unprecedented. Little is known, however, about the nuanced public opinions on the vaccines on social media.

\noindent\textbf{Methods:} We adopt a human-guided machine learning framework using more than six million tweets from almost two million unique Twitter users to capture public opinions on the vaccines for SARS-CoV-2, classifying them into three groups: pro-vaccine, vaccine-hesitant, and anti-vaccine. After feature inference and opinion mining, 10,945 unique Twitter users are included in the study population. Multinomial logistic regression and counterfactual analysis are conducted.

\noindent\textbf{Results:} Socioeconomically disadvantaged groups are more likely to hold polarized opinions on COVID-19 vaccines – either pro-vaccine ($B=0.40, SE = 0.08, p<.001, OR = 1.49; 95\% CI = [1.26, 1.75]$) or anti-vaccine ($B = 0.52, SE = 0.06, p<.001, OR = 1.69; 95\% CI = [1.49, 1.91]$). People who have the worst personal pandemic experience are more likely to hold the anti-vaccine opinion ($B = -0.18, SE = 0.04, p<.001, OR = 0.84; 95\% CI = [0.77, 0.90]$). The U.S. public is most concerned about the safety, effectiveness, and political issues regarding vaccines for COVID-19, and improving personal pandemic experience increases the vaccine acceptance level.

\noindent\textbf{Conclusion:} Opinion on COVID-19 vaccine uptake varies across people of different characteristics. 

\end{abstract}
\vspace{3mm}
\section{Introduction}

Researchers suggest that the transmission of SARS-CoV-2 will quickly rebound if interventions (e.g., quarantine and social distancing) are relaxed~\cite{ferguson2020report}. Vaccination has greatly reduced the burden of many infectious diseases~\cite{andre2008vaccination} throughout history, and developing SARS-CoV-2 vaccines that can be used globally is, therefore, a priority for ending the pandemic~\cite{yamey2020ensuring}. Nevertheless, as scientists and medical experts around the world are developing and testing COVID-19 vaccines, the U.S. public is now divided over whether or not to obtain COVID-19 vaccines. According to a recent Pew Research Center study\footnote{https://www.pewresearch.org/science/2020/09/17/u-s-public-now-divided-over-whether-to-get-covid-19-vaccine/ [Accessed July 20, 2021]}, in May, 2020 71\% of U.S. adults indicated that they would definitely or probably obtain a vaccine to prevent COVID-19 if it were available. The percentage dropped sharply, however, to 51\% in September, 2020. The survey shows that the U.S. public is concerned about the safety and effectiveness of possible vaccines, and the rapid pace of the approval process. 

Previous studies show that the sharing of public concerns about vaccines might lead to delaying or not getting vaccination~\cite{gust2008parents}, which could compromise global COVID-19 vaccine distribution strategies. This phenomenon is termed “vaccine hesitancy”~\cite{dube2013vaccine} which is a complex issue driven by a variety of context-specific factors~\cite{larson2014understanding}. Researchers have investigated public opinions on existing vaccines for vaccine-preventable diseases like MMR~\cite{motta2018knowing, deiner2019facebook}, HPV~\cite{abdelmutti2010risk, pan2020caught} and H1N1~\cite{henrich2011public}. Hesitancy and opinions can vary, however, according to the vaccine involved~\cite{bedford2007more}. 

\citet{lazarus2020global} and \citet{feleszko2020flattening} have investigated the potential acceptance of a COVID-19 vaccine using survey methods, yet little is known about the scope and causes of public opinions on COVID-19 vaccines on social media platforms. Although the survey data of a traditional design can lead to detecting causality, it is labor-intensive and expensive~\cite{mokdad2010measuring}, thus, being difficult to deploy surveys at a large scale without introducing social desirability biases~\cite{krumpal2013determinants} and in a timely manner, compared to social media data~\cite{mokdad2010measuring}. In addition, due to the passive nature of collecting social media data, observing social media data can potentially capture a different (and unperturbed) view of human behaviors~\cite{heikinheimo2017user}. To the best of our knowledge, there is no other study that has tracked and understood the public opinion regarding COVID-19 vaccines using social media data.

Meanwhile, the development and testing of COVID-19 vaccines has drawn great attention and response on social media platforms like Twitter and Reddit that allow fast sharing of health information~\cite{scanfeld2010dissemination,singh2020first, yeung2020face} and are found to play a major role in disseminating information about vaccinations~\cite{stahl2016impact,dunn2017mapping,brainard2020misinformation,tangcharoensathien2020framework, wu2021characterizing}. Public attitudes towards the vaccines, therefore, can be reflected by analyzing comments and posts in social media~\cite{kim2020effects, tomeny2017geographic}.

In the current study, we adopt a human-guided machine learning framework based on state-of-the-art transformer language models to capture individual opinions on COVID-19 vaccines, and categorize these opinions into three groups: pro-vaccine, vaccine-hesitant, anti-vaccine. We use more than 40,000 rigorously selected tweets (out of over six million tweets collected using keywords) posted by over 20,000 distinct Twitter users ranging from September to November of 2020. We aggregate the tweets to reflect the state-level and the national attitudes towards COVID-19 vaccines. To characterize the opinion groups, we extract and infer individual-level features such as demographics, social capital, income, religious status, family status, political affiliations, and geo-locations. \citet{lazarus2020global} suggested that personal experience such as COVID-19 sickness in the people and their family, and the external perception such as cases and mortality per million of a nation’s population are associated with the vaccine acceptance level. To quantitatively measure and confirm these two effects, we extract the sentiment of personal pandemic experience and non-pandemic experience for each Twitter user. We collect the number of COVID-19 daily confirmed cases from the data repository maintained by the Center for Systems Science and Engineering (CSSE) at Johns Hopkins University to measure the county-level pandemic severity perception. In our study, we hypothesize that:
\begin{itemize}
    \item \textbf{Hypothesis 1:} There will be differences in demographics, social capital, income, religious status, family status, political affiliations and geo-locations among opinion groups.
    \item \textbf{Hypothesis 2:} The personal pandemic experience will have an impact on shaping the attitude towards potential COVID-19 vaccines.
    \item \textbf{Hypothesis 3:} The county-level pandemic severity perception will have an impact on shaping the attitude towards potential COVID-19 vaccines.
\end{itemize}

We conduct multinomial logistic regression and find that there are differences in demographics, social capital, income, religious status, political affiliations and geo-locations among the opinion groups. People who have the worst personal pandemic experience are more likely to hold anti-vaccine opinion. In addition, people who have the worst pandemic severity perception are more likely to be vaccine-hesitant. We further show that the individual-level features can be used to anticipate whether this person is in favor of the potential COVID-19 vaccines - or not - over time. By incorporating the individual-level features and additional factor indicators, and by conducting counterfactual analyses, we find that the U.S. public is most concerned about the safety, effectiveness, and political issues with regard to potential vaccines for COVID-19 and improving personal pandemic experience increases the vaccine acceptance level. 

\section{Materials and Methods}

The Methods section is structured as follows. We describe the datasets we use in Methods M1 and how we infer or extract features in Methods M2. We describe our strategy for opinion mining and the standard of labelling in Methods M3. In Methods M4, we discuss the experimental procedures.

\subsection{M1 DataSets}
\subsubsection{Twitter}  We use the Tweepy API\footnote{https://www.tweepy.org/ [Accessed July 21, 2021]} to collect the related tweets which are publicly available. The search keywords and hashtags are COVID-19 vaccine-related or vaccine-related, including ``vaccine'', ``COVID-19 vaccine'', ``COVID vaccine'', ``COVID19 vaccine'', ``vaccinated'', ``immunization'', ``covidvaccine'', ``\#vaccine'' and ``covid19vaccine''. It is noteworthy that the capitalization of non-hastag keywords does not matter in the Tweepy query. Slang and misspellings of the related keywords are also included which are composed of ``vacinne'', ``vacine'', ``antivax'' and ``anti vax''. In the end, 6,314,327 tweets (including retweets) from September 28 to November 4, 2020 posted by 1,874,468 unique Twitter users are collected. To collect as many related text as possible, both COVID-19 vaccine-related and vaccine-related search keywords are used. However, the tweets collected using the vaccine-related search keywords are not necessarily related to COVID-19 vaccines. For example, MMR vaccine-related or HPV vaccine-related tweets might be crawled as well. In addition, the data collection is carried out during the flu shot season, resulting in collecting many influenza shot-related tweets. We apply a keyword-based search in tweets to remove all the tweets containing MMR, autism, HPV, tuberculosis, tetanus, hepatitis B, flu shot or flu vaccine (4.0\% removed).

The tweet content and other Twitter profile information are used to extract or predict demographics, user-level features like the number of {\tt followers}, income, religious status, family status, political affiliations, geo-locations, sentiment about the COVID-19-related experience and non-COVID-related experience. To infer the family status, religious status and sentiment, we use Tweepy API to collect the publicly available tweets posted by each user for the last three months. For example, if the tweet containing the search keywords or hashtags was posted on October 1, 2020, then all the publicly available tweets posted by this Twitter user from July 1 to October 1, 2020 are collected as well. It should be noted that only the last 3,200 tweets can be collected per the Tweepy API limitations.

\begin{figure*}[htbp]
    \centering
    \includegraphics[trim=150 0 150 0,clip,width=\linewidth]{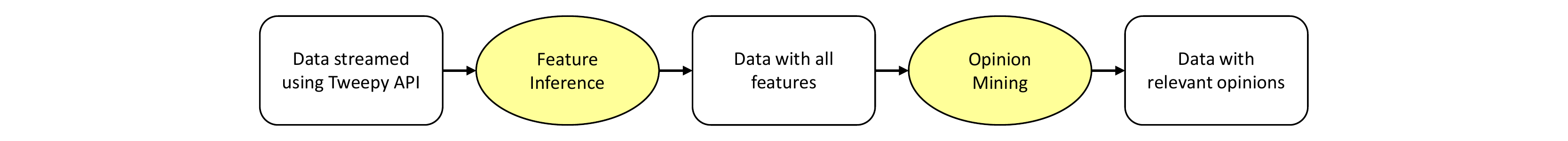}
    \caption{The diagram of data preprocessing procedures.}
    \label{fig:preprocess}
\end{figure*}

The preprocessing pipeline is shown in Figure~\ref{fig:preprocess}. First, the features of the Twitter users are inferred or extracted. To better understand the relationships between all characteristics, we choose to only keep the users of which we can infer all the features except for sentiment. Next, we achieve the mining of opinions via a human-guided machine learning framework. 25,407 unique users with all the features except for the sentiment scores are used to study the temporal and spatial patterns of the opinions. 10,945 of them with sentiment scores are further included in the characterization study and counterfactual analyses.

\subsubsection{JHU CSSE}
We extract the number of COVID-19 daily confirmed cases from the data repository maintained by the Center for Systems Science and Engineering (CSSE) at Johns Hopkins University~\cite{JHUcsse2019}. The median relative change of the number of daily confirmed cases of the last three months at the county level is calculated to measure the county-level pandemic severity perception.

\subsection{M2 Feature Inference}

\subsubsection{Demographics} Following the methods of \citet{lyu2020sense}, we use the Face++ API\footnote{https://www.faceplusplus.com/ [Accessed July 21, 2021]} to infer the gender and age information of the users using their profile images. The invalid image urls and images with multiple or zero faces are excluded. The gender and age information of the remaining users (i.e., there is only one intelligible face in the profile image) is inferred. Since our study focuses on the opinions of U.S. adults, the users who are younger than 18 are removed. Face++ can achieve a good accuracy of the gender and age inference of Twitter data~\cite{jung2018assessing}.

\subsubsection{User-level features}
Seven user-level features are crawled by Tweepy API as well which include the number of {\tt followers}, {\tt friends}, {\tt listed memberships}, {\tt favourites}, {\tt statuses}, the number of months since the user account was created, and the {\tt Verified} status. Moreover, we normalize the number of {\tt followers}, {\tt friends}, {\tt listed memberships}, {\tt favourites}, and {\tt statuses} by the number of months since the user account was created.

\subsubsection{Geo-locations}
For Twitter, we choose to resolve the geo-locations using users' profiles. Similar to \citet{lyu2020sense}, the locations with noise are excluded, and the rest are classified into urban, suburban, or rural.

\subsubsection{Income}
Following the method of \citet{preoctiuc2015studying}, we use a supervised ensemble model to predict the income of Twitter users. The ensemble model includes Gradient Boost Decision Tree (GBDT), Random Forest, Logistic Regression, and XGboost. We use the income datasets of Twitter users~\cite{preoctiuc2015studying} to train our model(s). The features include age, days of Twitter history, the number of {\tt followers}, {\tt friends}, {\tt listed memberships}, {\tt favourites}, and sentiment score calculated by Vader~\cite{gilbert2014vader}. We categorize income into three classes (low, medium, high) based on the income levels of \citet{kochhar2018american} and turn regression problems into classification problems. The accuracy of is $70.02\%$.

\subsubsection{Religious Status} We assign each user a boolean value for whether she/he is religious based on the tweets and the description in the profile~\cite{zhang2021influence}. 

\subsubsection{Family Status} By applying regular expression search, we identify users who show evidence that they are either fathers or mothers~\cite{zhang2021influence}. 

\subsubsection{Political Affiliations}
The political attribute is labelled based on whether this Twitter user followed the Twitter accounts of the top political leaders. The incumbent president (Joe Biden. Joe Biden was the presidential candidate when the data were collected.) and the former president (Donald Trump) are included in the analysis. Due to limitation of Twitter API, only about half of Donald Trump's follower ID was crawled.

\subsubsection{Sentiment} In our study, we intend to infer the sentiment of personal pandemic experience and non-pandemic experience. First, we use keyword search methods to classify the three-month historical tweets into COVID-related and non-COVID-related. If a  tweet does not contain any of the keywords:
``corona'', ``covid'', ``covid19'', ``coronavirus'', ``chinese virus'', ``china virus'' , ``wuhan virus'',``wfh'', ``work from home'', ``pandemic'', ``epidemic'', ``herd immunity'', ``quarantine'', ``lockdown'', ``mortality'', ``morbidity'', ``social distancing'', ``mask'', ``social distance'', ``respirator'', ``state of emergency'', ``ventilator'', ``isolation'' , ``fatality'', ``community spread'', ``vaccine'', ``vaccinated'', ``vaccination'', ``panic buying'' , ``hoard'',  it is categorized as non-COVID-related. The example tweets are 

``{\tt $<user>$ I can not wait to take the last name of my husband! I feel so good to solidify our union by taking his name.I also  cringe a little bit at the whole “keep the maiden name on social media” thing some girls do...I’m more “leave-and-cleave” type.}''

and

``{\tt $<user>$ what a distinguished day that was.}''

The remaining tweets are categorized as COVID-19-related. The example tweets are

``{\tt i am the type of person who does half an hour of meditation and yoga from my peloton app before going to bed to read some chapters of my book and be fast asleep before 11pm. quarantine changed me.}''

and 

``{\tt $<user>$ Oooorr...I can wear a mask, get on an plane, in a limited space, with NO social distancing, with people from hundreds of different households, ALL going to various destinations, and then take my mask OFF to eat/drink once I’m in my seat $<hashtag>$ $<hashtag>$}''

For each Twitter user, the tweets of the two categories are concatenated, respectively. Next, a normalized, weighted composite score is calculated to measure the sentiment of the tweet content using Vader~\cite{gilbert2014vader}. The score is between -1 (most extreme negative) and +1 (most extreme positive). Vader outperforms individual human raters when assessing the sentiment of tweets~\cite{gilbert2014vader}.

\subsection{M3 Opinion Mining}
To capture the opinions expressed through text by Twitter users, we adopt a human-guided machine learning framework inspired by \citet{sadilek2013nemesis}. The text are classified into four categories: (1) pro-vaccine, (2) vaccine-hesitant, (3) anti-vaccine, and (4) irrelevant.

Tweets might be retweeted for multiple times. We observe that there are 6,703 non-unique tweets in the initial batch of over 90,000 tweets. These non-unique tweets, combined with their retweets constitute 62.9\% of all tweets. As a result, the tweets are divided into two groups - the unique-tweet group and the non-unique-tweet group. 430 non-unique tweets which have been retweeted for at least 20 times are included in the non-unique-tweet group. These tweets and their retweets constitute 41.5\% of all tweets. The rest are included in the unique-tweet group. All the tweets of the non-unique-tweet group are manually annotated. However, only a subgroup of the unique-tweet group are manually annotated. The state-of-the-art transformer-based language model~\cite{yang2019xlnet}, trained with the subgroup, is used to make estimates of the rest of the unique-tweet group.

\subsubsection{Human-guided machine learning framework}
We annotate the opinions of the tweets as pro-vaccine, vaccine-hesitant, or anti-vaccine using a human-guided machine learning framework to strike the best balance between automation and accuracy. In total, we stream over six million publicly available tweets from Twitter using Tweepy API between September 28 to November 4, 2020 with search keywords that are vaccine-related or COVID-19 vaccine-related. Unlike~\citet{tomeny2017geographic}, a majority of the tweets crawled with the search keywords in our study is irrelevant to the actual individual opinions about the vaccines for COVID-19, which causes a challenging class imbalance problem that may not only slow down the annotation process but also hinder the performance of automated classifiers~\cite{japkowicz2002class}. To address this problem, we adopt a human-guided machine learning framework~\cite{sadilek2013nemesis} based on the state-of-the-art transformer language model to label the opinions of the tweets.  After extracting or inferring the features of these tweets and their authors, we only keep the ones with all the required informative features available.

We initialize the human-guided machine learning framework by sampling 2,000 unique tweets from the corpus $C$ with 244,049 tweets. Three researchers independently read each tweet and make a judgement whether this tweet is irrelevant, pro-vaccine, vaccine-hesitant, or anti-vaccine. Table~\ref{tab:label_standard} describes the labeling scheme for each opinion category. We label each tweet as one of the categories as long as it matches one of the descriptions of that category. The label of the tweet is assigned with the consensus votes from three researchers. If three researchers vote entirely differently, the senior researcher determines the label of this tweet after discussing with the other two researchers. The Fleiss' Kappa score of the three researchers is 0.52. The corpus $C_{train}$ of the initial 2,000 labelled tweets is fed to the XLNet model~\cite{yang2019xlnet}. The four-class classification model $H_{1}$ is trained and validated on an external validation set $D_{validation}$ with 400 annotated tweets. The distribution of the four categories is balanced. We then construct another binary classification model $H_{2}$ that is trained with only two classes of data. The data for $H_{1}$ and $H_{2}$ are almost the same except for the label of the output variable. For $H_{2}$, one class includes all the irrelevant tweets of the data for $H_{1}$ and the other includes all the relevant tweets that are  pro-vaccine, vaccine-hesitant, or anti-vaccine in the data for $H_{1}$. After training, $H_{2}$ is used to make estimates for a corpus of 4,500 unlabeled tweets sampled from $C$ regarding whether they are irrelevant or relevant. 90\% of a new batch of corpus is composed of the top 10\% of the most likely relevant tweets. The other 10\% of the new batch is sampled uniformly at random to increase diversity. This new batch of corpus of 500 tweets is annotated by the three researchers as aforementioned and is added to the corpus $C_{train}$. $H_{1}$ is trained with the updated $C_{train}$ and validated again. This whole process is considered as one iteration. For each iteration, the three researchers annotate a new batch of corpus of 500 tweets.

\begin{table*}[htbp]
    \centering
    \small
    \begin{tabular}{|c|l|}
    \hline
    \textbf{Category} &  
   \multicolumn{1}{|c|}{\textbf{Description}}    \\
    \hline
    \multirow{3}{7em}{Pro-vaccine} &	i.  Claiming that they would take the vaccine once it is available \\ 
    & ii.  Advocating and supporting vaccine/vaccine-associated entities like vaccine experiment trials \\ 
    & iii.  Believing that the vaccine will be the solution to the pandemic \\
    \hline
    \multirow{4}{7em}{Vaccine-hesitant} &	i.  Claiming that they would like to take the vaccine after the vaccine is proven safe/effective  \\ 
    & ii.  Claiming that they would wait for a while and see whether a vaccine is truly safe/effective \\
    &if there is one \\ 
    & iii.  Showing worries about the effectiveness of a rushed vaccine \\
    \hline
    \multirow{6}{7em}{Anti-vaccine} &	i.  Promoting/arguing in favor of conspiracy theory about vaccine/vaccine-associated entities  \\ 
    & ii.  Believing that an effective vaccine would not be invented quickly and help overcome \\ &the pandemic \\
    & iii.  Believing that a covid-19 vaccine is dangerous for whatever reasons and would not take it \\ & even though the commenters claim that they are not anti-vaccine \\
    \hline
    \multirow{4}{7em}{Irrelevant} 
    & i.  Vaccine News. No written opinion from the commenters  \\ 
    & ii.  Including vaccine and the commenters’ opinions, but the focus is something else \\ &  (i.e., insurance, politics, personal life experience, economics, emotional complaints, etc.) \\
    & iii.  Comments/questions on vaccines/vaccine-associated entities but with unclear meanings  \\
    \hline
    \end{tabular}
    \caption{Labeling scheme for Tweets.}
    \label{tab:label_standard}
\end{table*}

\begin{figure}[htbp!]
    \centering
    \includegraphics[width=\linewidth]{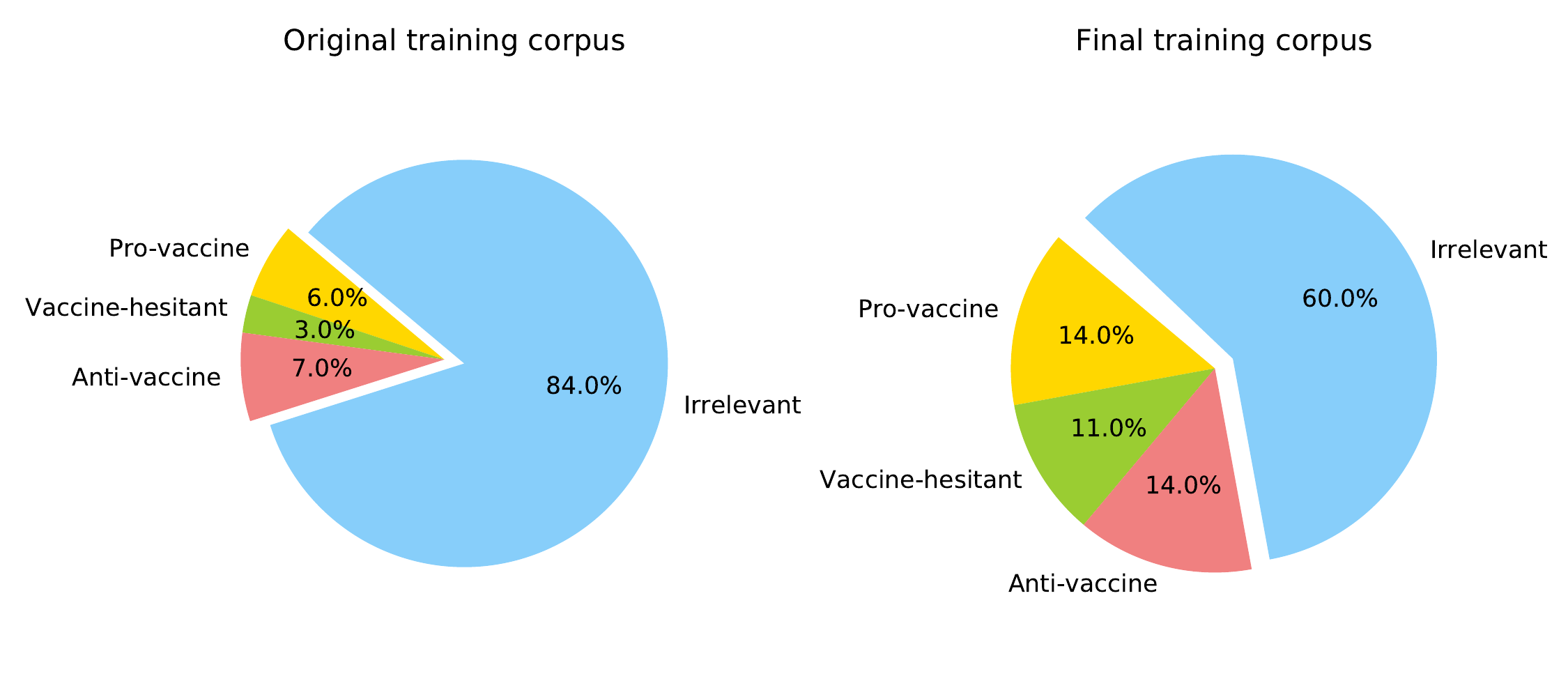}
    \caption{Distributions of different categories of the original and final training corpora.}
    \label{fig:dataset_distribution}
\end{figure}

\begin{figure}[htbp!]
    \centering
    \includegraphics[width=\linewidth]{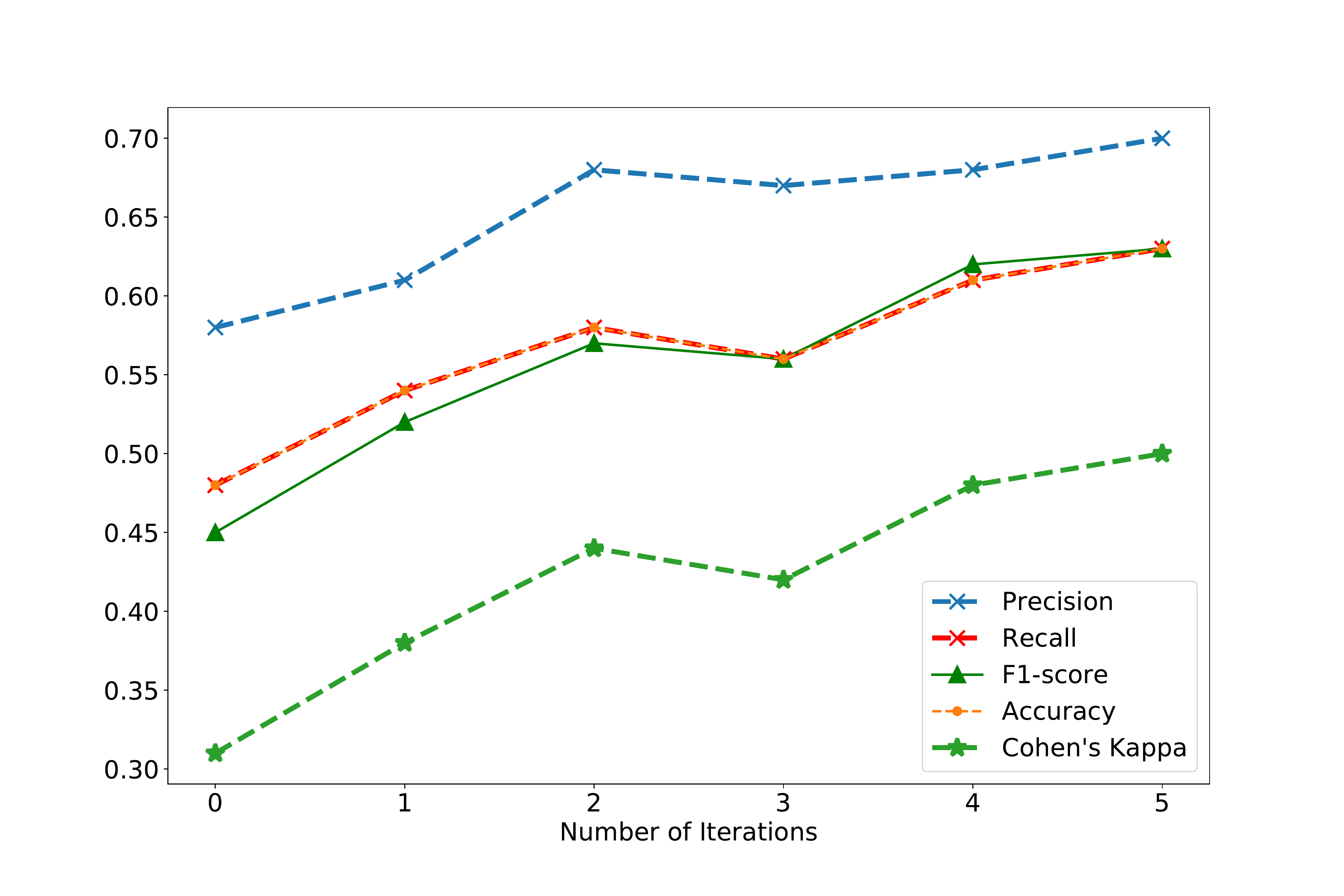}
    \caption{Performance of $H_{1}$ of each iteration.}
    \label{fig:xlnet_performance}
\end{figure}

This framework actively searches for relevant tweets to increase the sizes of the relevant datasets. Figure~\ref{fig:dataset_distribution} shows the percentages of the different opinion groups of the original $C_{train}$ and the final $C_{train}$ after five iterations. In each iteration, humans guide the machine to learn the irrelevant, pro-vaccine, vaccine-hesitant, and anti-vaccine tweets by updating the training set. Figure~\ref{fig:xlnet_performance} shows the performance of $H_{1}$ of each iteration. As a result, the framework allows us to label the opinions of the tweets and build the model more efficiently.

\subsubsection{Tweets preprocessing} We adopt a tweet preprocessing pipeline from \citet{baziotis-etal-2017-datastories-semeval} which can transform the specific text often used in Twitter to special tokens. For example, if the original tweet is 

``{\tt Scientists develop a COVID vaccine that could initiate a 10-times stronger immune response $<url>$ }''

After preprocessing, the tweet becomes

``{\tt scientists develop a $<allcaps>$ covid $</allcaps>$ vaccine that could initiate a $<number>$ - times stronger immune response $<url>$ }'' 

\subsubsection{Performance of the XLNet model} Table~\ref{tab:xlnet_result_tab} summarizes the performance of the final four-class XLNet model $H_{1}$ on the external validation set with 400 samples. The final accuracy is 0.63 and the Cohen's Kappa score is 0.5, which indicates a good agreement. 

\begin{table}[htbp]
    \centering
   
    \begin{tabular}{|c|c|c|c|}
    \hline
    \textbf{Class}&\textbf{Precision} & \textbf{Recall} & \textbf{F1-score} \\
    \hline
    Irrelevant & 0.45&0.84&0.59\\
    Pro-vaccine & 0.78& 0.52&0.62\\
    Vaccine-hesitant & 0.77 & 0.54 & 0.64\\
    Anti-vaccine & 0.79 & 0.61 & 0.69\\
    \hline
    Overall & 0.7 & 0.63 & 0.63\\
    \hline
    \end{tabular}
     \caption{Performance of the four-class XLNet model $H_{1}$.}
    \label{tab:xlnet_result_tab}
\end{table}

\subsection{M4 Analysis Details}
\subsubsection{Statistical Analysis}
To understand what opinion (i.e., pro-vaccine, vaccine-hesitant, and anti-vaccine) the people ($n=10,945$) would hold based on the their demographics, social capital, income, religious status, family status, political affiliations, geo-location, sentiment about COVID-19-related experience and non-COVID-related experience, and relative change of the number of daily confirmed cases at the county level, we conduct multinomial logistic regression, selecting vaccine-hesitant group as the reference category.

\subsubsection{Counterfactual analyses}
Following \citet{chang2020mobility}, we intend to estimate the impact of communication-related strategies by constructing a hypothetical machine learning model that reflects the expected effect. To assess the potential outcomes of the communication-related strategies, we build the machine learning model using the real data, and apply the constructed model to the hypothetical data.

\begin{figure}[htbp]
    \centering
    \includegraphics[width = \linewidth]{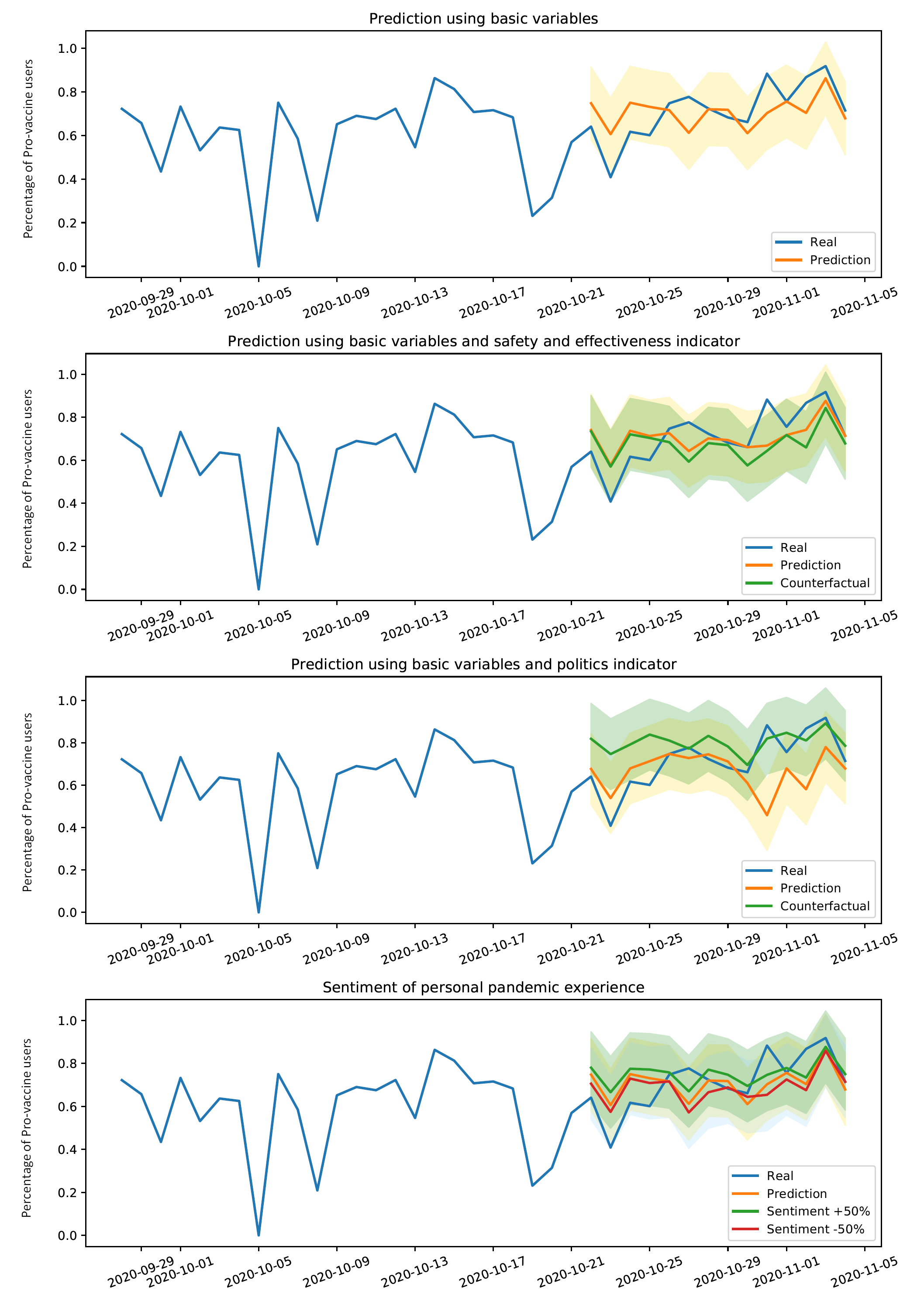}
    \caption{Counterfactual analyses illustrate the importance of politics, safety and effectiveness factor indicators, and personal pandemic experience.}
    \label{fig:counter}
\end{figure}

The data ranging from September 28 to October 21, 2020 are used to train a support vector machine (SVM) $H_{3}$ which makes predictions about the opinion group of the data of the latest two weeks (October 22 - November 4, 2020). The real percentage of pro-vaccine users and the prediction percentage are plotted in Figure~\ref{fig:counter}. The real percentage falls within in one standard deviation of the predicted percentage, indicating a good simulation performance.

\begin{figure*}
    \centering
    \includegraphics[width = \textwidth]{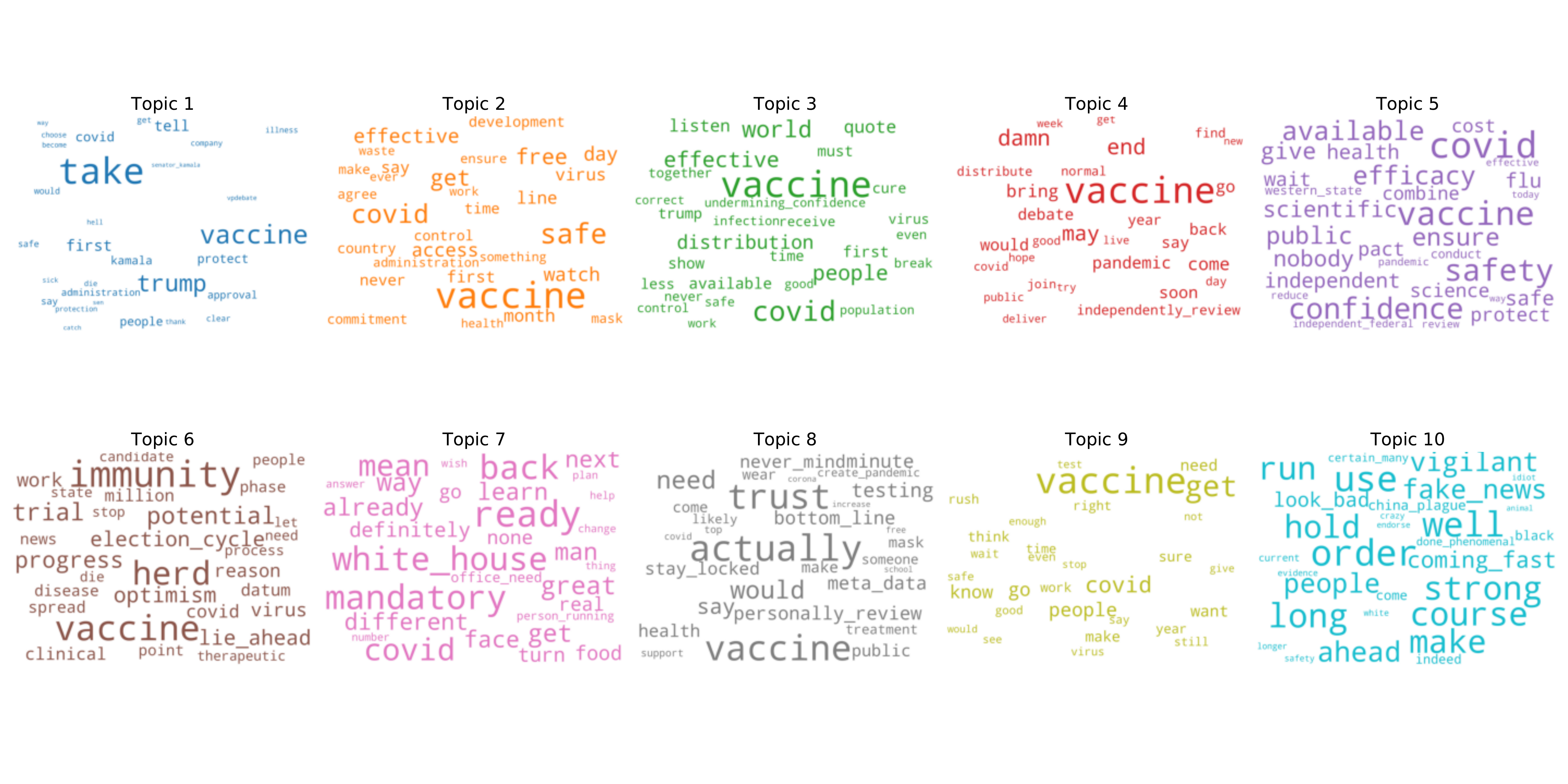}
    \caption{10 topics extracted from the tweets with the top 30 keywords.}
    \label{fig:topic}
\end{figure*}

We further analyze the relationship between the opinions and the topics of the tweets using the Latent dirichlet allocation (LDA) topic modelling~\cite{blei2003latent} with 10 topics as shown in Figure~\ref{fig:topic}. The coherence score is 0.31. In the word cloud of each topic, top 30 keywords are plotted. As we can see from the figure, people are most concerned about the safety and effectiveness of the vaccine which is consistent with the Pew Research Center survey\footnote{https://www.pewresearch.org/science/2020/09/17/u-s-public-now-divided-over-whether-to-get-covid-19-vaccine/ [Accessed July 20, 2021]}. Some politics-related keywords like ``administration'', ``white house'', and the names of political figures like ``Trump'' and ``Kamala'' are presented as well. To label the factor indicators, we narrow down the 10 topics to two major ones: ``safety and effectiveness'' and ``politics'', and use keyword search methods. The keywords for the safety and effectiveness include ``safe'', ``effective'', and ``efficacy''. The keywords for the politics include ``administration'', ``politics'', ``politician'', ``political'' and the names of Donald Trump, Mike Pence, Joe Biden and Kamala Harris. Each tweet is labelled 1 if it contains the related keywords, and 0 if it does not. Table~\ref{tab:factor} shows the descriptive statistics of these two variables. The basic settings for the counterfactual classifiers are the same as $H_{3}$. We analyze one factor at a time. We train the classifier with the basic variables and the factor indicator with the real value. The basic variables include user demographics, Twitter usage patterns, sentiment of the pandemic and non-pandemic experience, income, religious status, family status, political affiliation, as well as the population density. The prediction is plotted in orange in Figure~\ref{fig:counter}. Then we change the value of the factor indicator which was originally 1 into 0, keeping other variables constant. The trained classifier is applied to the hypothetical data, and the prediction is plotted in green.

\begin{table}[htbp]
    \centering
    \scriptsize
    \begin{tabular}{l c c c c c}
    \hline
     \textbf{Variables}    & \textbf{N} &\textbf{Mean} & \textbf{SD} &\textbf{Min} & \textbf{Max} \\
     \hline
       1. Politics  & 10,945 & 0.2512 & 0.4337 & 0 & 1\\
       2. Safety and effectiveness  & 10,945 & 0.1801 & 0.3843 & 0 & 1\\
       \hline
    \end{tabular}
    \caption{Descriptive statistics of the factor indicators.}
    \label{tab:factor}
\end{table}



\section{Results}

\subsection{Characterization of different opinion groups}
\begin{figure*}[htbp]
     \centering

     \begin{subfigure}[b]{\textwidth}
         \centering
         \includegraphics[width=\textwidth]{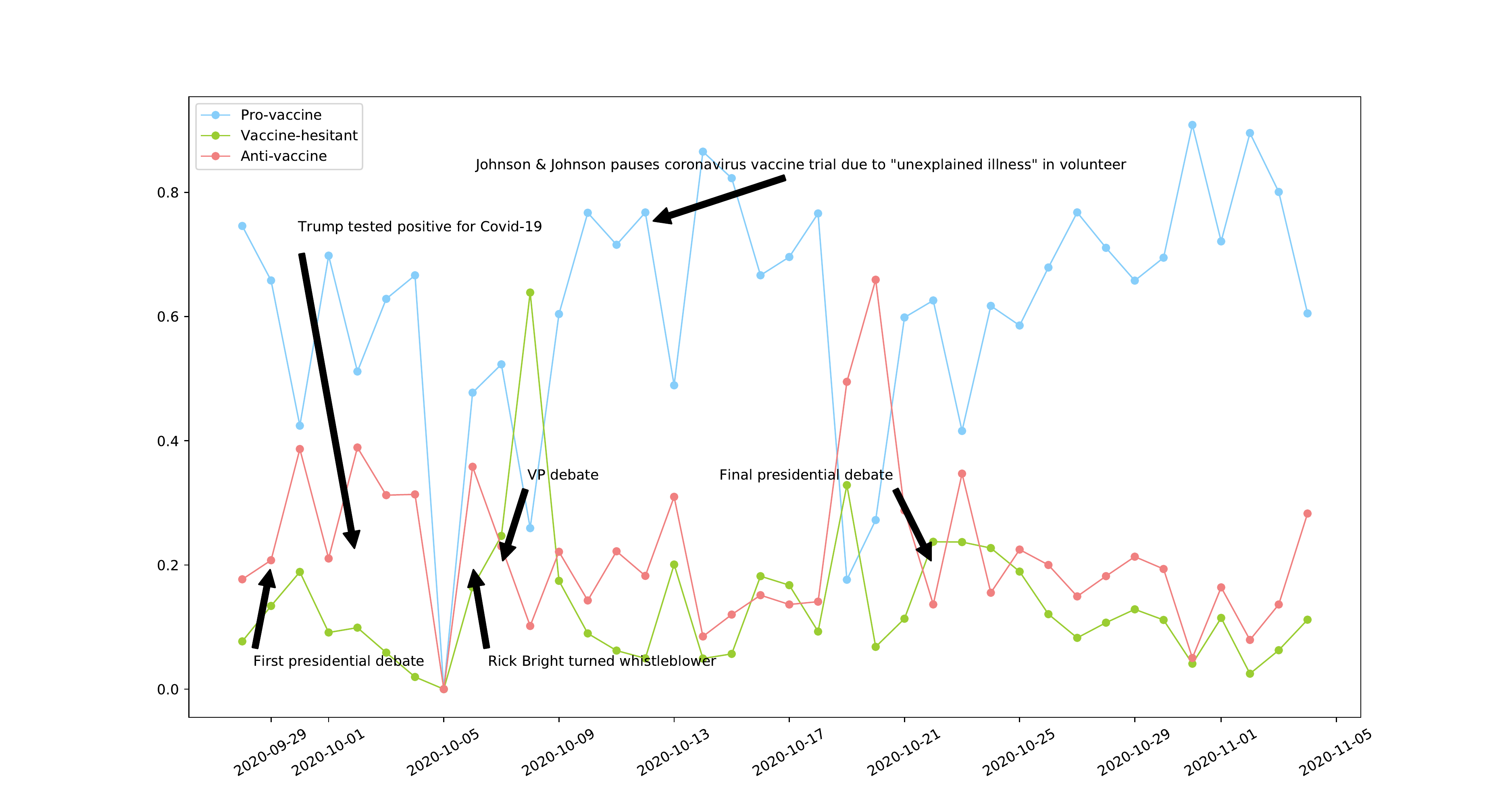}
         \caption{}
         \label{fig:us_level}
     \end{subfigure}
     \hfill
     \begin{subfigure}[b]{\textwidth}
         \centering
         \includegraphics[width=\textwidth]{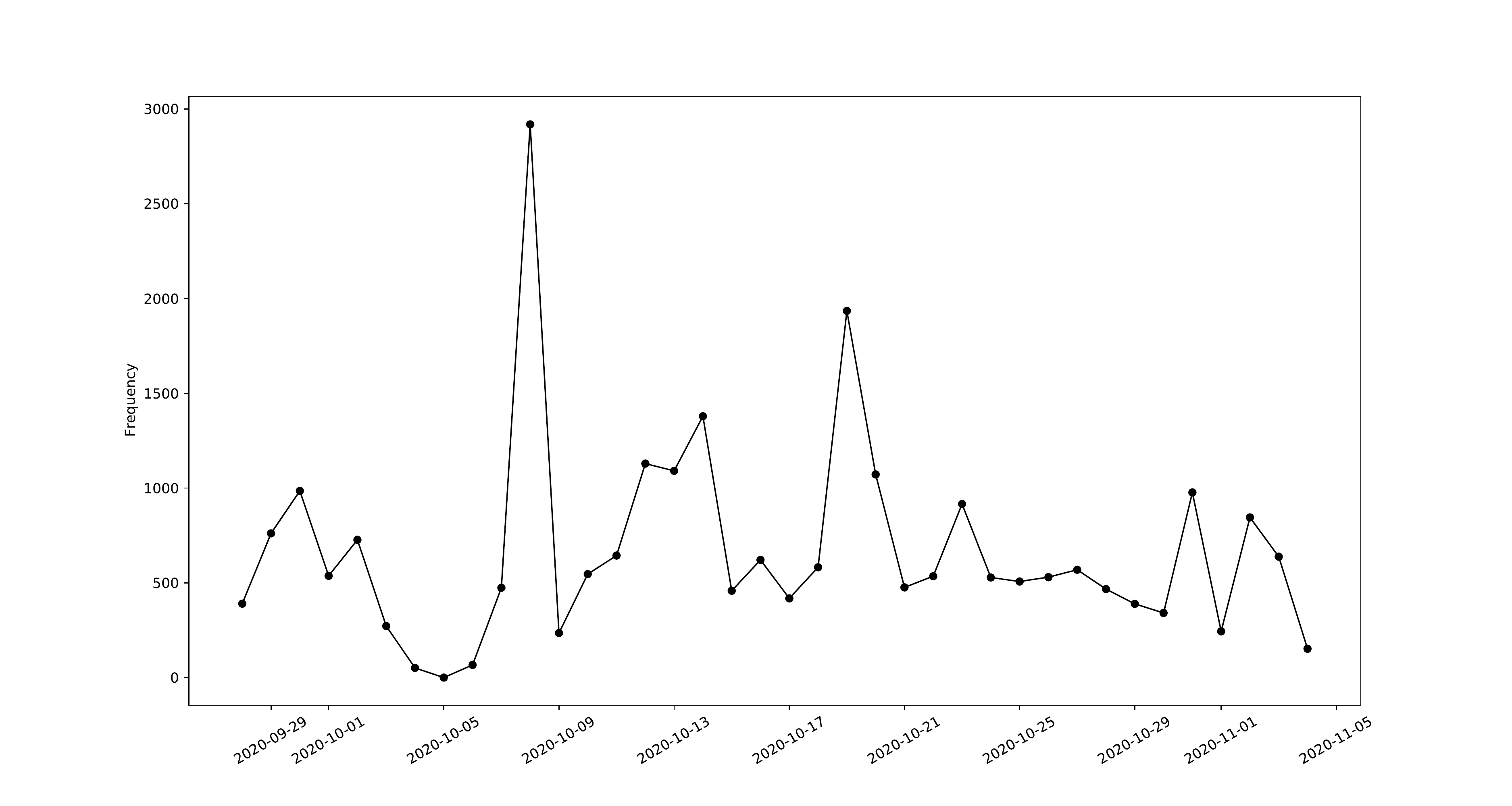}
         \caption{}
         \label{fig:abs_number}
     \end{subfigure}
     
        \caption{(a) The proportions of the opinion groups from September 28 to November 4, 2020.  (b) Number of Twitter users from September 28 to November 4, 2020. The data of October 5, 2020 are missing due to a data collection issue.}
        \label{fig:temporal_spatial}
\end{figure*}

\begin{figure*}[htbp]
    \centering
    \includegraphics[width = \textwidth]{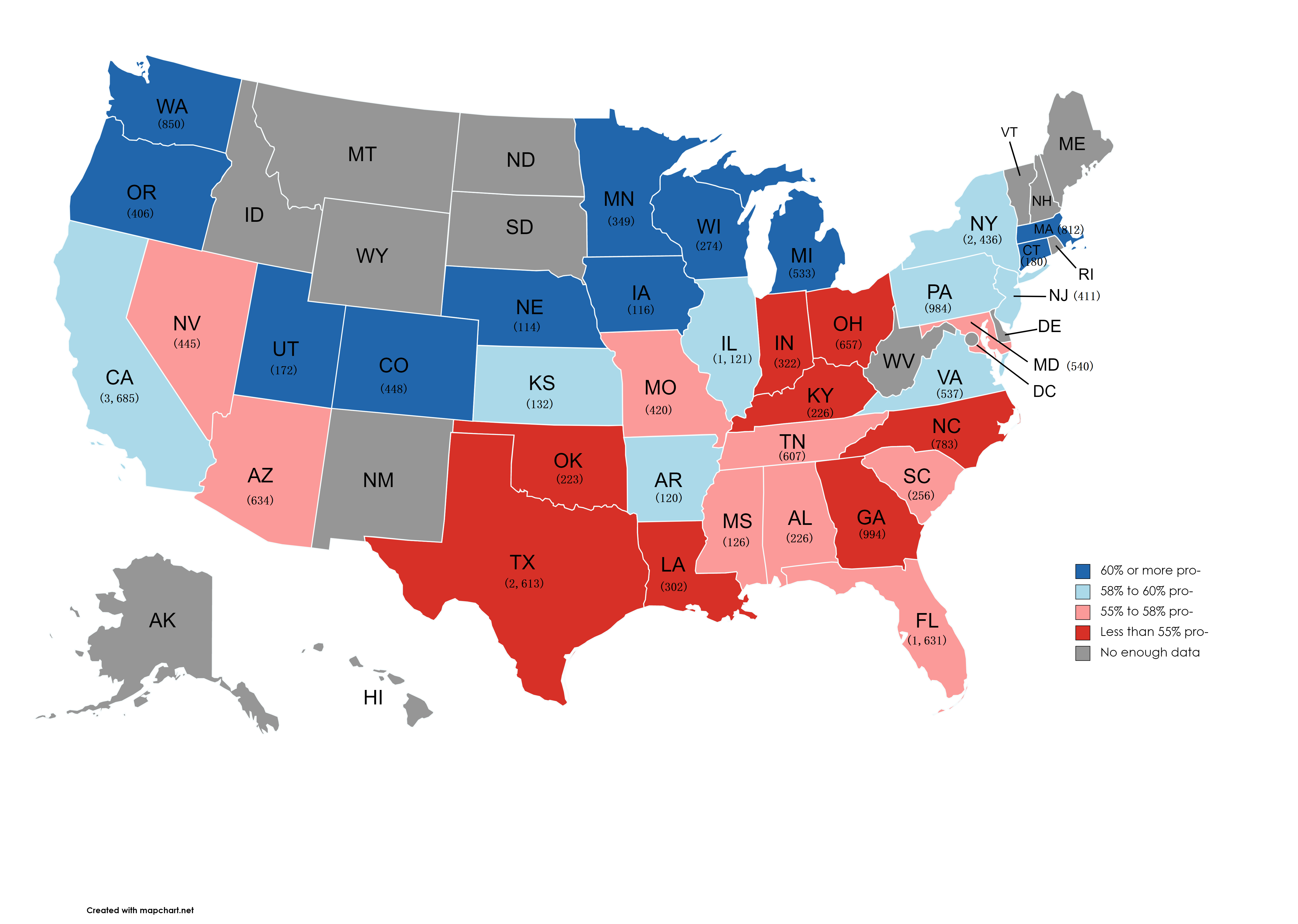}
    \caption{State-level public opinions about potential COVID-19 vaccines. The numbers in parentheses stand for the sizes of the study populations.}
    \label{fig:state_opinion}
\end{figure*}

The proportions of the different opinion groups of the U.S public change over time as shown in Figure~\ref{fig:us_level}, which roughly correspond to the major pandemic-related events. Figure~\ref{fig:abs_number} shows the number of Twitter users. Overall, 57.65\% (6,218 of 25,407) are pro-vaccine, 19.30\% (2,469 of 25,407) are vaccine-hesitant, and the rest are anti-vaccine. By aggregating people at the state level, we estimate the opinions about the potential COVID-19 vaccines of each state as shown in Figure~\ref{fig:state_opinion}. The Southeast of the U.S. shows a relatively lower acceptance level, so does the cluster of Ohio, Indiana and Kentucky.


\begin{figure*}
    \centering
    \includegraphics[width = \textwidth]{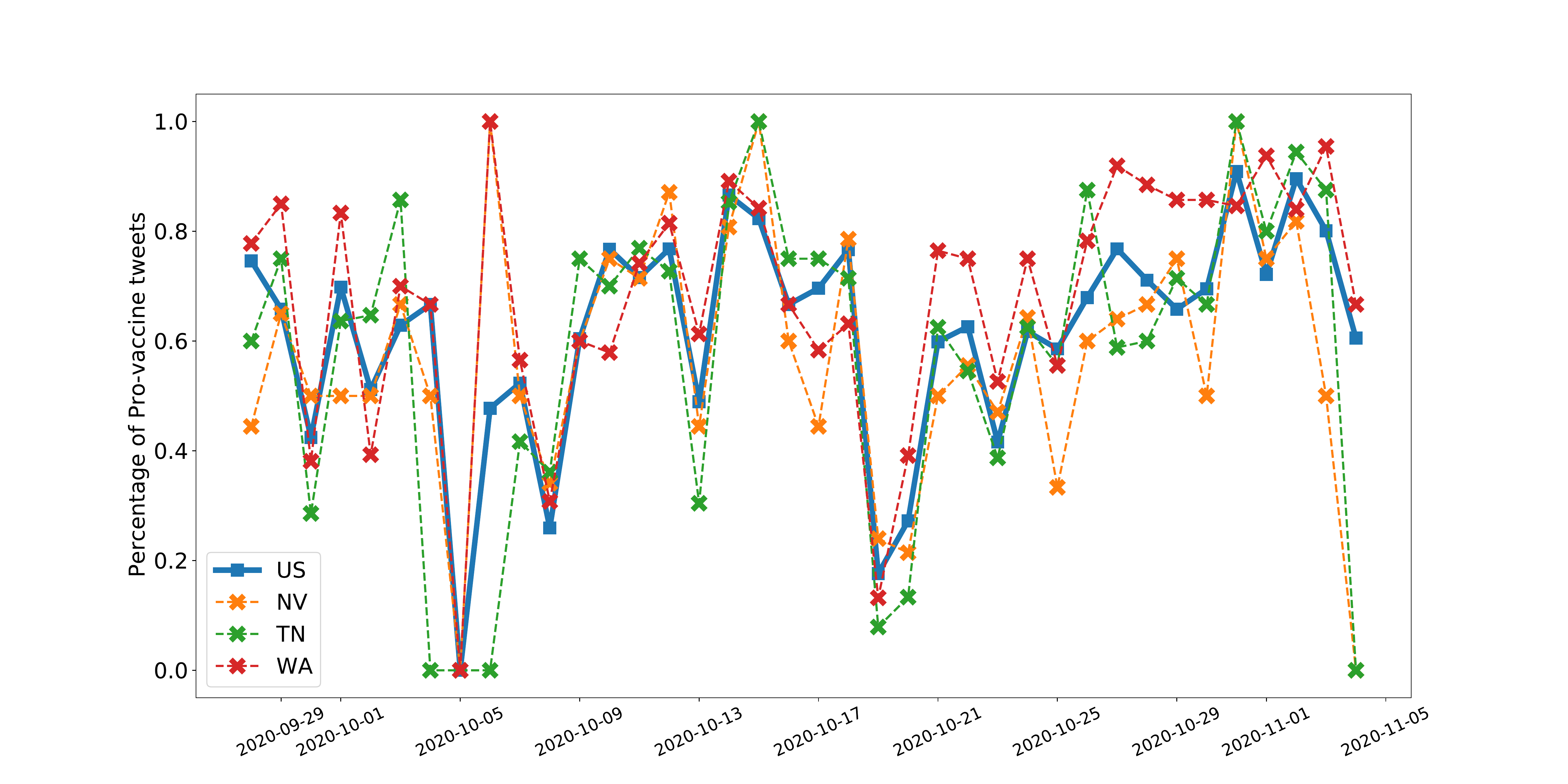}
    \caption{The percentages of the pro-vaccine groups of the national average, Nevada, Tennessee, and Washington.}
    \label{fig:granger}
\end{figure*}

After performing the Granger Causality Test with a one-day lag, we find that (Figure~\ref{fig:granger}), in Nevada, Tennessee and Washington, the percentage of the pro-vaccine people deviates the most from the national average ($p>.05$). The percentage of the pro-vaccine group of Washington is above the national average during the most of the time, while the acceptance level of Nevada is relatively lower than the national average. More drastic changes are observed for the acceptance level of Tennessee. 

Descriptive statistics and bi-variate correlations of the variables of the multinomial logistic regression are shown in Table~\ref{tab:char_desc}. Table~\ref{tab:regression_outputs} summarizes the results of the multinomial logistic regression. The Chi-square test shows that the variables significantly predict the opinion on potential COVID-19 vaccines: $\chi^{2}(40, N = 10,945)=1,340.94, p<.001$, McFadden's pseudo $R^{2}=.06$, which supports our hypotheses. Next, we show the predictive effects of these variables with paired comparisons.

\begin{sidewaystable*}[htbp]
    \centering
    
    \tabcolsep=0.1cm
    \scriptsize
    \begin{tabular}{l r r l l l l l l l l l l l l l l l l l l l l l l l}
    
    \hline
    \textbf{Variables} & \multicolumn{1}{c}{\textbf{Mean}} & \multicolumn{1}{c}{\textbf{SD}} & \multicolumn{1}{c}{\textbf{1}} &  \multicolumn{1}{c}{\textbf{2}} & \multicolumn{1}{c}{\textbf{3}} &\multicolumn{1}{c}{\textbf{4}} & \multicolumn{1}{c}{\textbf{5}} &\multicolumn{1}{c}{\textbf{6} }&\multicolumn{1}{c}{\textbf{7}} &\multicolumn{1}{c}{\textbf{8}} & \multicolumn{1}{c}{\textbf{9}} &\multicolumn{1}{c}{\textbf{10}} & \multicolumn{1}{c}{\textbf{11}} & \multicolumn{1}{c}{\textbf{12}} &\multicolumn{1}{c}{\textbf{13}}&\multicolumn{1}{c}{\textbf{14}}&\multicolumn{1}{c}{\textbf{15}}&\multicolumn{1}{c}{\textbf{16}}&\multicolumn{1}{c}{\textbf{17}}&\multicolumn{1}{c}{\textbf{18}}&\multicolumn{1}{c}{\textbf{19}}\\
    \hline
    1. Gender (0 = male, 1 = female) & 0.46& 0.50 \\
    2. Age (years) & 39.89 & 14.69 & -$.08^{**}$\\
    
    3. {\tt Verified} (0 = no, 1 = yes) &0.04 & 0.20 & -.00 & $.03^{**}$ \\
    4. Twitter history (months) & 91.30 & 43.47 & -.02 & $.03^{**}$ & $.09^{**}$\\
    5. \# {\tt Followers} & 1.60 & 1.63 & $.04^{**}$ & -.01  & $.37^{**}$ & -$.08^{**}$\\
    6. \# {\tt Friends} & 1.95 & 1.25 & $.04^{**}$ & -.00  & $.09^{**}$ & -$.29^{**}$ & $.68^{**}$\\
    7. \# {\tt Listed memberships} & -1.62 & 0.93 & -.00 & $.08^{**}$  & $.49^{**}$ & $.22^{**}$ & $.69^{**}$ & $.31^{**}$\\
    8. \# {\tt Favorites} & 4.17 & 1.95 & $.12^{**}$ & -$.14^{**}$  & -$.03^{**}$ & -$.21^{**}$ & $.38^{**}$ & $.47^{**}$ & $.05^{**}$\\
    9. \# {\tt Statuses} & 4.09 &1.43& $.02^{*}$ & -$.11^{**}$  & $.07^{**}$ & -$.09^{**}$ & $.53^{**}$ & $.45^{**}$ & $.29^{**}$ & $.58^{**}$\\
    10. Higher-Income (0 = no, 1 = yes) & 0.00 & 0.05 & -.00 & -$.03^{**}$  & $.03^{**}$ & -.01 & $.05^{**}$ & $.02^{*}$ & $.04^{**}$ & $.02^{**}$ & .01\\
    11. Lower-Income (0 = no, 1 = yes) & 0.76 & 0.43 & -.00 & -$.43^{**}$  & -$.04^{**}$ & $.05^{**}$ & -$.15^{**}$ & -$.18^{**}$ & -$.12^{**}$ & -$.13^{**}$ & -$.09^{**}$ & -$.10^{**}$\\
    12. Religious (0 = no, 1 = yes) & 0.04 & 0.19 & .01 & $.07^{**}$  & -$.03^{**}$ & -$.03^{**}$ & $.03^{**}$ & $.07^{**}$ & -$.03^{**}$ & .02 & .02 & -.00 & -$.06^{**}$\\
    13. Having kids (0 = no, 1= yes) &0.12 & 0.32 & $.09^{**}$ & $.09^{**}$  & $.03^{**}$ & $.02^{*}$ & $.04^{**}$ & $.05^{**}$ & $.04^{**}$ & .01 & -$.05^{**}$ & -.01 & -$.05^{**}$ & $.09^{**}$\\
    14. Following Trump (0 = no, 1 = yes)& 0.11 & 0.31 & -$.05^{**}$ & $.06^{**}$ & -$.03^{**}$ & -$.17^{**}$ & -$.05^{**}$ & $.04^{**}$ & -$.12^{**}$ & -.01 & -$.05^{**}$ & -.00 & -$.05^{**}$ & $.06^{**}$ & $.04^{**}$\\
    15. Following Biden (0 = no, 1 = yes)& 0.17 & 0.38 & $.09^{**}$ & $.07^{**}$ & $.06^{**}$ &  $.10^{**}$ & $.05^{**}$ & $.19^{**}$ & $.06^{**}$ & $.09^{**}$ & $.03^{**}$ & .00 & -$.06^{**}$ & -$.03^{**}$ & $.07^{**}$ & .01\\
    16. Rural (0 = no, 1 = yes)& 0.19 & 0.40 & -.02 & $.09^{**}$ & -$.05^{**}$ & -$.04^{**}$ & -$.07^{**}$ & -.02 & -$.08^{**}$ & -.01 & -$.04^{**}$ & .00 & -$.05^{**}$ & $.06^{**}$ & $.04^{**}$ & $.05^{**}$ & -.00 \\
    17. Suburban (0 = no, 1 = yes)& 0.14 & 0.35 & -.01 & $.07^{**}$  & -.01 & -$.02^{*}$ & -$.05^{**}$ & -$.03^{**}$ & -$.05^{**}$ & -$.03^{**}$ & -$.02^{**}$ & -.02 & -.01 &$.03^{**}$ & $.04^{**}$ & $.03^{**}$ & -.00 & -$.20^{**}$ \\
    \makecell[l]{18. Pandemic experience \\ (sentiment)} & 0.06 & 0.80 & $.03^{**}$ & -$.04^{**}$  & $.10^{**}$ & $.03^{**}$ & $.09^{**}$ & -.00 & $.15^{**}$ & -$.08^{**}$ & -$.08^{**}$ & $.02^{*}$ & $.04^{**}$ & -$.03^{**}$ & -.00 & -$.06^{**}$ & .01 & -$.03^{**}$ & .00\\
    \makecell[l]{19. Non-pandemic experience \\ (sentiment)} & 0.62 & 0.75 & $.07^{**}$ & -$.06^{**}$ & $.07^{**}$ & $.04^{**}$ & $.14^{**}$ & $.09^{**}$ & $.14^{**}$ & $.06^{**}$ & .01 & .02 & .02 & .00 & $.03^{**}$ & -$.06^{**}$ & .01 & -$.03^{**}$ & -$.02^{*}$ & $.27^{**}$\\
     \makecell[l]{20. Pandemic severity perception \\ (relative change of \# daily confirmed cases)}& 0.01&0.00& -.01& $.03^{**}$ & -$.03^{**}$ & -$.02^{*}$ & -$.05^{*}$ & -.01 & -$.07^{**}$ & -.00 & -$.04^{**}$ & .01 & -$.03^{**}$ & $.06^{**}$ & $.05^{**}$ & $.04^{**}$ &-.01 & $.27^{**}$ & $.12^{**}$ & -.01& .01\\
\hline
    \end{tabular}
    {\raggedright Note. * $p<0.05$. ** $p<0.01$. \par}
    \caption{Descriptive statistics and the bi-variate correlations. The numbers of {\tt followers}, {\tt friends}, {\tt listed memberships}, {\tt favorites}, {\tt statuses} are normalized by the months of Twitter history and log-transformed.}
    \label{tab:char_desc}
\end{sidewaystable*}

\begin{table*}[]
    \centering
    \small
    \setlength{\tabcolsep}{3pt} 
    \renewcommand{\arraystretch}{1.5} 
    \begin{tabular}{r l l l l l l}
    \hline
     & \multicolumn{3}{l}{\textbf{Anti-vaccine}} & \multicolumn{3}{l}{\textbf{Pro-vaccine}}  \\
     \cline{2-7}
     
     Predictor & \textit{B} & \textit{SE} & OR (95\% CI) & \textit{B} & \textit{SE} & OR (95\% CI) \\
\hline
Intercept &-$1.82^{***}$ &0.26 & & $0.79^{***}$ &0.20 & \\
     Age (years) & 0.00 & 0.00 & 1.00 (1.00, 1.01)&  $0.01^{***}$ &0.00  & 1.01 (1.01, 1.02) \\
Twitter history (months) & 0.00 &0.00& 1.00 (1.00, 1.00)& -$0.003^{***}$& 0.001 & 0.997 (0.996, 0.999)\\
           \# {\tt Followers} & $0.28^{***}$  & 0.04 & 1.32 (1.22, 1.42)& $0.08^{**}$ &0.03& 1.09 (1.02, 1.16)\\
           \# {\tt Friends} & -$0.18^{***}$ & 0.04 & 0.83 (0.77, 0.90) & -$0.07^{*}$ & 0.03 & 0.93 (0.88, 0.99)\\
           \# {\tt Listed memberships} & -$0.63^{***}$ & 0.06 & 0.53 (0.47, 0.60)& $0.10^{*}$ & 0.05 & 1.10 (1.01, 1.20)\\
           \# {\tt Favorites} & $0.04^{*}$& 0.03 & 1.12 (1.06, 1.19)& $0.04^{*}$& 0.02 & 1.04 (1.01, 1.08)\\
           \# {\tt Statuses} & $0.11^{***}$ & 0.03 & 1.12 (1.06, 1.19)& -$0.06^{*}$& 0.02& 0.94 (0.90, 0.99)\\
           Pandemic experience (sentiment) & -$0.18^{***}$ & 0.04 & 0.84 (0.77, 0.90)& $0.21^{***}$& 0.03& 1.24 (1.16, 1.32)\\
           Non-pandemic experience (sentiment) & -0.04 & 0.04 &0.96 (0.89, 1.04)& $0.13^{***}$& 0.03& 1.14 (1.07, 1.22)\\
           \makecell[r]{Pandemic severity perception \\ (relative change of \# daily confirmed cases)} & -14.99 & 8.13 & 0.00 (0.00, 2.58)& -$22.68^{***}$ & 6.59 & 0.00 (0.00, 0.00)\\
           Female & -$0.25^{***}$& 0.06 & 0.78 (0.69, 0.88)& -$0.47^{***}$& 0.05& 0.63 (0.57, 0.69)\\
           {\tt Verified} user & -$0.61^{*}$&0.27 & 0.54 (0.32, 0.91)& -0.16& 0.14& 0.85 (0.65, 1.12)\\
           Higher-income & -170.67&5.00e+36 & 0.00 (0.00, Inf)& 0.47 & 0.43 & 1.60 (0.70, 3.68)\\
           Lower-income & $0.40^{***}$& 0.08& 1.49 (1.26, 1.75)&$0.52^{***}$ & 0.06 &  1.69 (1.49, 1.91)\\
           Religious &$0.74^{***}$ &0.17 & 2.10 (1.52, 2.91)&$0.37^{*}$ &0.15& 1.45 (1.07, 1.95)\\
           Having kids & -0.11& 0.10 & 0.90 (0.74, 1.09)& 0.00& 0.08&1.00 (0.86, 1.15)\\
           Following Trump & $0.41^{***}$& 0.10 &1.51 (1.26, 1.83)& 0.06& 0.08&1.06 (0.90, 1.25)\\
           Following Biden & -$1.22^{***}$& 0.10 &0.29 (0.24, 0.36)&-$0.34^{***}$ & 0.06&0.71 (0.63, 0.80)\\
           Rural & $0.17^{*}$& 0.08 & 1.19 (1.01, 1.39)& 0.07 & 0.07 & 1.07 (0.94, 1.22)\\
           Suburban & $0.18^{*}$& 0.09 & 1.20 (1.01, 1.43)& 0.11 & 0.07 &1.12 (0.97, 1.29)\\
          \hline
          Chi-square & \multicolumn{6}{c}{$1,340.94^{***}$}\\
          \textit{df} & \multicolumn{6}{c}{40}\\
          $-2$ log likelihood & \multicolumn{6}{c}{20,171.90} \\
          McFadden's pseudo $R^{2}$ & \multicolumn{6}{c}{0.06} \\
        Sample size & \multicolumn{6}{c}{10,945}\\          
          \hline
          {\raggedright Note. * $p<0.05$. ** $p<0.01$. *** $p<0.001$.\par}
    \end{tabular}
    \caption{Multinomial logistic regression outputs for the opinion on potential COVID-19 vaccines against demographics and other variables of interest. The vaccine-hesitant group is selected as the reference category.}
    \label{tab:regression_outputs}
\end{table*}

\subsubsection{Women are more likely to hold hesitant opinions.}
Gender is statistically significant ($\chi^{2}=91.83, p<.001$). Women are likely to hold hesitant opinions rather than polarized opinions (i.e., pro-vaccine, anti-vaccine). Specifically, comparing the anti-vaccine group and vaccine-hesitant group, we find that women are less likely to be anti-vaccine ($B=-0.25, SE=0.06, p<.001, OR = 0.78; 95\% CI = [0.69, 0.88]$). Comparing the pro-vaccine group and vaccine-hesitant group, we find that women are also less likely to be pro-vaccine ($B = -0.47, SE = 0.05, p<.001, OR = 0.63; 95\% CI = [0.57, 0.69]$).

\subsubsection{Older people tend to be pro-vaccine.}
Age is statistically significant ($\chi^{2}=72.47, p<.001$). Comparing the anti-vaccine group and vaccine-hesitant group, we do not find significant evidence that older people are more anti-vaccine. However, comparing the pro-vaccine group and vaccine-hesitant group, we find that people who are one year older are 1.01 ($B=0.01, SE = 0.00 ,p<.001, OR=1.01; 95\% CI = [1.01, 1.02]$) times more likely to be pro-vaccine instead of vaccine-hesitant, which echoes the findings of \citet{lazarus2020global}. One potential explanation is that the risk of dying with COVID-19 increases with age~\citep{lloyd2020bearing}, and the benefits of not getting infected with COVID-19 outweigh the risk of getting vaccinated.

\subsubsection{Different patterns of Twitter usage.}
A {\tt Verified} Twitter account must represent or other wise be associated with a prominently recognized individual or brand\footnote{https://help.twitter.com/en/managing-your-account/about-twitter-verified-accounts [Accessed July 21, 2021]}. In our study, {\tt Verified} status is statistically significant ($\chi^{2}=6.12, p<.05$). Comparing the anti-vaccine group and vaccine-hesitant group, we find {\tt Verified} users are less likely to be anti-vaccine ($B = -0.61, SE = 0.27, p<.05, OR = 0.54; 95\% CI = [0.32, 0.91]$), however, comparing the pro-vaccine group and vaccine-hesitant group, we do not find significant differences. 

Months of Twitter history is statistically significant ($\chi^{2}=17.52, p<.001$). Comparing the anti-vaccine group and vaccine-hesitant group, we do not find significant differences, however, comparing the pro-vaccine group and vaccine-hesitant group, we find if the months of Twitter history were to increase by one month, it is 0.997 ($B=-0.003, SE = 0.001, p <.001, OR =0.997; 95\% CI = [0.996, 0.999]$) less likely to be pro-vaccine than vaccine-hesitant.

After normalizing the number of {\tt followers}, {\tt friends}, {\tt listed memberships}, {\tt favorites}, and {\tt statuses} with the number of months of Twitter history, we still find that the social capital is statistically significant. Specifically, there are significant differences in terms of {\tt followers} counts ($\chi^{2}=51.06, p<.001$), {\tt friends} counts ($\chi^{2}=21.28, p<.001$), {\tt listed memberships} counts ($\chi^{2}=199.51, p<.001$), {\tt favorites} counts ($\chi^{2}=6.10, p<.05$), {\tt statuses} counts ($\chi^{2}=47.37, p<.001$). 

Comparing the anti-vaccine group and vaccine-hesitant group, if the log-scale {\tt followers} count were to increase by one unit, it is 1.32 ($B=0.28, SE = 0.04, p<.001, OR = 1.32; 95\% CI = [1.22, 1.42]$) times more likely to be anti-vaccine. If the log-scale {\tt friends} count were to increase by one unit, it is less likely to be anti-vaccine ($B=-0.18, SE = 0.04, p<.001, OR = 0.83; 95\% CI = [0.77, 0.90]$). If the log-scale {\tt listed memberships} count were to increase by one unit, it is less likely to be anti-vaccine ($B=-0.63, SE = 0.06,p<.001, OR = 0.53; 95\% CI = [0.47, 0.60]$). If the log-scale {\tt favorites} count were to increase by one unit, it is 1.04 ($B=0.04, SE = 0.02, p<.05, OR = 1.04; 95\% CI = [1.00, 1.09]$) times more likely to be anti-vaccine. If the log-scale {\tt statuses} count were to increase by one unit, it is 1.12 ($B=0.11, SE = 0.03, p<.001, OR = 1.12; 95\% CI = [1.06, 1.19]$) times more likely to be anti-vaccine. 

Comparing the pro-vaccine group and vaccine-hesitant group, if the log-scale {\tt followers} count were to increase by one unit, it is 1.09 ($B=0.08, SE = 0.03, p<.01, OR = 1.09; 95\% CI = [1.02, 1.16]$) times more likely to be pro-vaccine. If the log-scale {\tt friends} count were to increase by one unit, it is less likely to be pro-vaccine ($B=-0.07, SE = 0.03, p<.05, OR = 0.93; 95\% CI = [0.88, 0.99]$). If the log-scale {\tt listed memberships} count were to increase by one unit, it is 1.11 ($B=0.10, SE = 0.05, p<.05, OR =1.10; 95\% CI = [1.01, 1.20]$) times more likely to be pro-vaccine. If the log-scale {\tt favorites} count were to increase by one unit, it is 1.04 ($B = 0.04, SE = 0.02, p<.05, OR = 1.04; 95\% CI = [1.01, 1.08]$) times more likely to be pro-vaccine. If the log-scale {\tt statuses} count were to increase by one unit, it is less likely to be pro-vaccine ($B = -0.06, SE = 0.02, p<.05, OR = 0.94; 95\% CI = [0.90, 0.99]$). 

Twitter users who have more {\tt followers} or fewer {\tt friends}, or give more {\tt favourites} are more likely to hold polarized opinion. The larger {\tt listed memberships} count is, the more likely the Twitter user is pro-vaccine. Twitter users who post more {\tt statuses} tend to be anti-vaccine.

\subsubsection{The lower-income group is more likely to hold polarized opinions.} 
Income is statistically significant ($\chi^{2}=79.09, p<.001$). Comparing the anti-vaccine group and vaccine-hesitant group, we find that the lower-income group is 1.49 ($B=0.40, SE = 0.08, p<.001, OR = 1.49; 95\% CI = [1.26, 1.75]$) times more likely to be anti-vaccine than the medium-income group, however,the difference between the higher-income group and medium-income group is not significant. 
Comparing the pro-vaccine group and vaccine-hesitant group, we find that lower-income group is 1.69 ($B = 0.52, SE = 0.06, p<.001, OR = 1.69; 95\% CI = [1.49, 1.91]$) times more likely to be pro-vaccine than medium-income group. The difference between the higher-income group and medium-income group is not significant. {\it Inconsistent} with \citet{lazarus2020global} that the higher the income is, the more likely people are pro-vaccine, we find the effect of income more nuanced. Lower-income people tend to be polarized. 

\subsubsection{Religious people are more likely to be polarized.} 
Religious status is statistically significant ($\chi^{2}=21.34, p<.001$). Comparing the anti-vaccine group and vaccine-hesitant group, we find that religious people are more likely to be anti-vaccine than non-religious people ($B = 0.74, SE = 0.17, p<.001, OR = 2.10; 95\% CI = [1.52, 2.91]$). Comparing the pro-vaccine group and vaccine-hesitant group, we find that religious people are also more likely to be pro-vaccine than non-religious people ($B = 0.37, SE = 0.15, p<.05, OR = 1.45; 95\% CI = [1.07, 1.95]$). This is in line with \citet{larson2014understanding} that the effect of religious status is complicated.

\subsubsection{Political diversion indicates a divided opinion about the potential COVID-19 vaccines.} Following Donald Trump is statistically significant ($\chi^{2}=25.22, p<.001$). Comparing the anti-vaccine group and vaccine-hesitant group, we find that the Twitter users who follow Donald Trump are 1.51 ($B = 0.41, SE = 0.10, p<.001, OR = 1.51; 95\% CI = [1.26, 1.83]$) times more like to be anti-vaccine than the Twitter users who do not. Comparing the pro-vaccine group and vaccine-hesitant group, following Donald Trump is not significant.

Following Joe Biden is statistically significant ($\chi^{2}=177.96, p<.001$). Comparing the anti-vaccine group and vaccine-hesitant group, we find that the Twitter users who follow Joe Biden are less like to be anti-vaccine than the Twitter users who do not ($B = -1.22, SE = 0.10, p<.001, OR = 0.30; 95\% CI = [0.24, 0.36]$). Comparing the pro-vaccine group and vaccine-hesitant group, we find that the Twitter users who follow Joe Biden are also less likely to be pro-vaccine than the Twitter users who do not ($B = -0.34, SE = 0.06, p<.001, OR = 0.71; 95\% CI = [0.63, 0.80]$).

Twitter users who follow Donald Trump tend to be anti-vaccine, while those who follow Joe Biden tend to be vaccine-hesitant.

\subsubsection{People living in suburban or rural areas are more likely to be anti-vaccine.} Although the population density of the area is not statistically significant across three opinion categories, we still find differences between the anti-vaccine group and vaccine-hesitant group. People living in suburban areas are 1.20 ($B = 0.18, SE = 0.09, p<.05, OR = 1.20; 95\% CI = [1.01, 1.43]$) times more likely to be anti-vaccine than people living in urban areas. People living in rural areas are 1.20 ($B = 0.17, SE = 0.08, p<.05, OR = 1.18; 95\% CI = [1.01, 1.39]$) times more likely to be anti-vaccine than people living in urban areas. 

Most of the results are consistent with Hypothesis 1. There are significant differences in demographics, social capital, income, religious status, political affiliations and geo-locations among opinion groups, however, we do not find significant difference in family status.

\subsubsection{Personal experience with COVID-19 and the county-level pandemic severity perception shape the opinion.} The sentiment score of personal experience with COVID-19 is statistically significant ($\chi^{2}=146.50, p<.001$). Comparing the anti-vaccine group and vaccine-hesitant group, we find that if the sentiment score of personal experience with COVID-19 were to increase by one unit (i.e., the sentiment became more positive), the person would be less likely to hold anti-vaccine opinion ($B = -0.18, SE = 0.04, p<.001, OR = 0.84; 95\% CI = [0.77, 0.90]$). Comparing the pro-vaccine group and vaccine-hesitant group, we find if the sentiment score of personal experience with COVID-19 were to increase by one unit (i.e., the sentiment became more positive), the person would be 1.24 times more likely to hold pro-vaccine opinion ($B = 0.21, SE = 0.03, p<.001, OR = 1.24; 95\% CI = [1.16, 1.32]$), which is consistent with Hypothesis 2.

The sentiment score of non-COVID-related personal experience is overall statistically significant ($\chi^{2}=29.28, p<.001$), but comparing the anti-vaccine group and vaccine-hesitant group, we find no significant difference. However, comparing the pro-vaccine group and vaccine-hesitant group, we find if the sentiment score of non-COVID-related personal experience were to increase by one unit (i.e., the sentiment became more positive),  the person would be more likely to hold pro-vaccine opinion ($B = 0.13, SE = 0.03, p<.001, OR = 1.14; 95\% CI = [1.07, 1.22]$).

The county-level pandemic severity perceptions are overall statistically significant ($\chi^{2}=11.76, p<.01$), supporting Hypothesis 3, but we find no significant difference comparing the anti-vaccine group and vaccine-hesitant group. However, comparing the pro-vaccine group and vaccine-hesitant group, if the relative change of the number of daily confirmed cases at the county level were to increased by one unit, the person would be less likely to be pro-vaccine ($B = -22.68, SE = 6.59, p<.001, OR = 0.00; 95\% CI = [0.00, 0.00]$).

At the individual level, the personal pandemic experience is a strong predictor of the opinion about COVID-19 vaccines. People who have the worst personal pandemic experience are more likely to hold anti-vaccine opinion. However, the non-pandemic experience is not a strong predictor of anti-vaccine opinion. At the county level, people who have the worst pandemic severity perception (i.e., the relative change of the number of daily confirmed cases is the largest) are more likely to be vaccine-hesitant.

\subsection{Counterfactual analyses}
\subsubsection{Removing the safety and effectiveness factors reduces the vaccine acceptance level. However, removing the politics factor increases it.} Figure~\ref{fig:counter} shows the results of counterfactual analyses of factor indicators. Using counterfactual analysis by turning the factor indicator of safety and effectiveness into 0, there is a clear decrease (4.42\% on average) of the percentage of the pro-vaccine people. However, by turning the factor indicator of politics into 0, there is a clear increase (22.65\% on average) of the percentage of the pro-vaccine people. This indicates that people are most concerned about the relationship between the politics and the potential COVID-19 vaccines, which is also mirrored by the news report\footnote{https://www.nytimes.com/2020/08/02/us/politics/coronavirus-vaccine.html [Accessed July 21, 2021]}.

\subsubsection{Improving personal pandemic experience increases the vaccine acceptance level.} Figure~\ref{fig:counter} shows the results of counterfactual analyses of different sentiment levels of personal pandemic experience. By increasing the sentiment scores with a factor of 50\%, the percentage of the pro-vaccine people increases by 6.39\%. However, by reducing the sentiment scores of a factor of 50\%, the percentage of the pro-vaccine people decreases by 2.82\%.

\subsection{Robustness verification}
The multinomial logistic regression and counterfactual analyses are conducted based on the opinion derived from the social media data. By adopting the human-guided machine learning framework, the dataset is composed of human-annotated data and machine-inferred data. Although the Cohen's Kappa score of the machine and human annotators reaches 0.5 after five iterations, the results of the analyses could potentially change given the seemingly limited prediction performance. Therefore, we additionally verify the robustness of the findings of multinomial logistic regression and counterfactual analyses by examining two different combinations of the human-annotated data and machine-inferred data:
\begin{itemize}
    \item Only human-annotated data (n = 2,939) are used to conduct the multinomial logistic regression and counterfactual analyses
    \item Human-annotated data (n = 2,939) and 50\% machine-inferred data (randomly sampled, n = 4,003) are used to conduct the multinomial logistic regression and counterfactual analyses
\end{itemize}
The results of the two robustness verification experiments are shown in Appendices. Both of them are basically consistent with the reported results in Table~\ref{tab:regression_outputs} and Figure~\ref{fig:counter}. The findings do not fundamentally change, confirming that the opinion computationally derived from the social media data are reliable for our study.

\section{Discussion}
We conduct multinomial logistic regression to investigate the scope and causes of public opinions on vaccines and test three hypotheses. The current study shows the hypothesized effects of most of the characteristics in predicting the odds of being pro-vaccine or anti-vaccine against vaccine-hesitant. The findings suggest that women are more vaccine-hesitant, which is consistent with the Reuters/Ipsos survey\footnote{https://www.reuters.com/business/healthcare-pharmaceuticals/more-women-than-men-us-nervous-about-fast-rollout-covid-vaccine-thats-problem-2020-12-11/ [Accessed July 21, 2021]}, and older people tend to be pro-vaccine. With respect to social capital, people who have more {\tt followers} or fewer {\tt friends}, or give more {\tt favorites}, are more likely to hold polarized opinions. {\tt Verified} status, months of Twitter history, {\tt listed memberships} counts and {\tt statuses} counts are statistically significant as well. We also show that the lower-income group is more likely to hold polarized opinions. This is {\it inconsistent} with the finding by \citet{lazarus2020global}. Moreover, religious people tend to hold polarized opinions. As for political affiliations, Twitter users who follow Donald Trump are more likely to be anti-vaccine rather than vaccine-hesitant, while those who follow Joe Biden tend to be vaccine-hesitant rather than anti-vaccine or pro-vaccine. In addition, we find people who live in rural or suburban areas tend to be anti-vaccine. However, we do not find the hypothesized predictive effect of family status on the opinion about vaccines.

Furthermore, the current study shows the hypothesized predictive effects of the personal pandemic experience and the county-level pandemic severity perception. In particular, personal experience with COVID-19 is a strong predictor of anti-vaccine opinion. The more negative the experience is, the more negative the opinion on vaccines is. People are more likely to be vaccine-hesitant if their pandemic severity perceptions are worse.

Our current study has limitations. The public opinions of some (less populated) states cannot be reflected due to the inadequate data. The findings could be further validated in other populations. In addition, the performance of the XLNet model of this study could be improved so that the proposed human-guided machine learning framework could identify the public opinions on the COVID-19 vaccines more accurately. Specifically, future work could address the class imbalance issue by annotating more tweets or augmenting the training data via back translation (i.e. translating English tweets to another language and then translating them back to English). Despite the limitations, our study broadly captures the public opinions on the potential vaccines for COVID-19 on Twitter. The final F1-score of our study is 0.63, indicating a good performance, which is achieved via the human-guided machine learning framework that is composed of both the state-of-the-art transformer-based language model and the large annotated dataset. By aggregating the opinions, we find a lower acceptance level in the Southeast part of the U.S. The changes of the proportions of different opinion groups correspond roughly to the major pandemic-related events. We show the hypothesized predictive effects of the characteristics of the people in predicting pro-vaccine, vaccine-hesitant, and anti-vaccine group. 
Using counterfactual analyses, we find that people are most concerned about the safety, effectiveness and politics regarding potential COVID-19 vaccines, and improving personal experience with COVID-19 increases the vaccine acceptance level. 

Our results can guide and support policymakers making more effective distribution policies and strategies. First, more efforts of dissemination should be spent on the socioeconomically disadvantaged groups who are exposed to potentially higher risks~\cite{chang2020mobility,hopman2020managing, adams2020inequality} and already possess more polarized attitudes towards the vaccines. Second, messaging for the vaccines is extremely important because the vaccine acceptance level can be increased by removing the politics factor. Third, safety and effectiveness issues need to be well addressed because the acceptance level is reduced by removing this factor. Finally, improving personal pandemic experience may increase the vaccine acceptance level as well and thus all helpful measures should be integrated to maximize the vaccine acceptance. In the future, by combining social media data and more traditional survey data, we hope to acquire deeper insights into the public opinions on potential COVID-19 vaccines and thus inform more effective vaccine dissemination policies and strategies.

\bibliography{main}

\appendix

\section{Supplementary Materials}
\subsection{Descriptive statistics of the variables}
\subsubsection{Age.}
Table~\ref{tab:age_des} shows the descriptive statistics of age.

\begin{table}[htbp]
\centering
\scriptsize
\begin{tabular}{llllll}
\hline
Group            & Mean  & SD    & Median & Minimum & Maximum \\
\hline
Pro-vaccine      & 40.85 & 14.64 & 39.00  & 18.00   & 96.00   \\
Vaccine-hesitant & 39.52 & 14.92 & 36.00  & 18.00   & 86.00   \\
Anti-vaccine     & 37.62 & 14.29 & 34.00  & 18.00   & 90.00  \\
\hline
\end{tabular}
\caption{Descriptive statistics of age.}
\label{tab:age_des}
\end{table}

\subsubsection{Characteristics of the Twitter users of three opinion groups.}
Table~\ref{tab:des_pro_vaccine}, Table~\ref{tab:des_vaccine_hesitant}, and Table~\ref{tab:des_anti_vaccine} show the descriptive statistics of the characteristics of the Twitter users of pro-vaccine, vaccine-hesitant, and anti-vaccine groups, respectively.

\begin{table}[]
\centering
\small
\begin{tabular}{lll}
\hline
Variables                               & Mean  & SD    \\
\hline
1. Gender (0 = male, 1 = female)        & 0.42  & 0.49  \\
2. Age (years)                          & 40.85 & 14.64 \\
3. {\tt Verified} (0 =no, 1 = yes)            & 0.05  & 0.22  \\
4. Twitter history (months)             & 99.54 & 43.34 \\
5. {\tt \# Followers}                         & 1.55  & 1.68  \\
6. {\tt \# Friends}                           & 1.83  & 1.24  \\
7. {\tt \# Listed memberships}                & -1.56 & 0.98  \\
8. {\tt \# Favourites}                        & 3.95  & 2.04  \\
9. {\tt \# Statuses}                          & 3.89  & 1.41  \\
10. Higher-Income (0 = no, 1 = yes)     & 0.00  & 0.06  \\
11. Lower-Income (0 = no, 1 = yes)      & 0.77  & 0.42  \\
12. Religious (0 = no, 1 = yes)         & 0.03  & 0.18  \\
13. Having kids (0 = no, 1 = yes)       & 0.12  & 0.33  \\
14. Following Trump (0 = no, 1= yes)    & 0.10  & 0.30  \\
15. Following Biden (0 = no, 1 = yes)   & 0.18  & 0.38  \\
16. Rural (0 = no, 1 = yes)             & 0.19  & 0.39  \\
17. Suburban (0 = no, 1 = yes)          & 0.14  & 0.35  \\
18. Pandemic experience (sentiment)     & 0.16  & 0.80  \\
19. Non-pandemic experience (sentiment) & 0.68  & 0.70  \\
20. Pandemic severity perception        & 0.01  & 0.00 \\
\hline
\end{tabular}
\caption{Descriptive statistics of the characteristics of the Twitter users of the pro-vaccine group. The numbers of {\tt followers}, {\tt friends}, {\tt listed memberships}, {\tt favorites}, {\tt statuses} are normalized by the months of Twitter history and log-transformed.}
\label{tab:des_pro_vaccine}
\end{table}

\begin{table}[]
\centering
\small
\begin{tabular}{lll}
\hline
Variables                               & Mean  & SD    \\
\hline
1. Gender (0 = male, 1 = female)        & 0.54  & 0.50  \\
2. Age (years)                          & 39.52 & 14.91 \\
3. {\tt Verified} (0 =no, 1 = yes)            & 0.04  & 0.20  \\
4. Twitter history (months)             & 102.83 & 42.89 \\
5. {\tt \# Followers}                         & 1.40  & 1.57  \\
6. {\tt \# Friends}                           & 1.85  & 1.16  \\
7. {\tt \# Listed memberships}                & -1.66 & 0.89  \\
8. {\tt \# Favourites}                        & 4.03  & 1.79  \\
9. {\tt \# Statuses}                          & 3.94  & 1.38  \\
10. Higher-Income (0 = no, 1 = yes)     & 0.00  & 0.06  \\
11. Lower-Income (0 = no, 1 = yes)      & 0.71  & 0.45  \\
12. Religious (0 = no, 1 = yes)         & 0.02  & 0.15  \\
13. Having kids (0 = no, 1 = yes)       & 0.12  & 0.33  \\
14. Following Trump (0 = no, 1= yes)    & 0.10  & 0.30  \\
15. Following Biden (0 = no, 1 = yes)   & 0.25  & 0.43  \\
16. Rural (0 = no, 1 = yes)             & 0.19  & 0.39  \\
17. Suburban (0 = no, 1 = yes)          & 0.13  & 0.34  \\
18. Pandemic experience (sentiment)     & 0.00  & 0.79  \\
19. Non-pandemic experience (sentiment) & 0.58  & 0.78  \\
20. Pandemic severity perception        & 0.01  & 0.00 \\
\hline
\end{tabular}
\caption{Descriptive statistics of the characteristics of the Twitter users of the vaccine-hesitant group. The numbers of {\tt followers}, {\tt friends}, {\tt listed memberships}, {\tt favorites}, {\tt statuses} are normalized by the months of Twitter history and log-transformed.}
\label{tab:des_vaccine_hesitant}
\end{table}

\begin{table}[]
\centering
\small
\begin{tabular}{lll}
\hline
Variables                               & Mean  & SD    \\
\hline
1. Gender (0 = male, 1 = female)        & 0.47  & 0.50  \\
2. Age (years)                          & 37.62 & 14.29 \\
3. {\tt Verified} (0 =no, 1 = yes)            & 0.01  & 0.10  \\
4. Twitter history (months)             & 92.95 & 43.84 \\
5. {\tt \# Followers}                         & 1.39  & 1.45  \\
6. {\tt \# Friends}                           & 1.77  & 1.15  \\
7. {\tt \# Listed memberships}                & -1.91 & 0.64  \\
8. {\tt \# Favourites}                        & 4.30  & 1.70  \\
9. {\tt \# Statuses}                          & 4.18  & 1.45  \\
10. Higher-Income (0 = no, 1 = yes)     & 0.00  & 0.00  \\
11. Lower-Income (0 = no, 1 = yes)      & 0.79  & 0.41  \\
12. Religious (0 = no, 1 = yes)         & 0.06  & 0.23  \\
13. Having kids (0 = no, 1 = yes)       & 0.10  & 0.30  \\
14. Following Trump (0 = no, 1= yes)    & 0.15  & 0.36  \\
15. Following Biden (0 = no, 1 = yes)   & 0.08  & 0.27  \\
16. Rural (0 = no, 1 = yes)             & 0.21  & 0.41  \\
17. Suburban (0 = no, 1 = yes)          & 0.15  & 0.36  \\
18. Pandemic experience (sentiment)     & -0.16  & 0.75  \\
19. Non-pandemic experience (sentiment) & 0.50  & 0.84  \\
20. Pandemic severity perception        & 0.01  & 0.00 \\
\hline
\end{tabular}
\caption{Descriptive statistics of the characteristics of the Twitter users of the anti-vaccine group. The numbers of {\tt followers}, {\tt friends}, {\tt listed memberships}, {\tt favorites}, {\tt statuses} are normalized by the months of Twitter history and log-transformed.}
\label{tab:des_anti_vaccine}
\end{table}

\subsection{Robustness verification}
We verify the robustness of the findings of multinomial logistic regression and counterfactual analyses by examining two different combinations of the human-annotated data and machine-inferred data:
\begin{itemize}
    \item Only human-annotated data are used to conduct the multinomial logistic regression and counterfactual analyses
    \item Human-annotated data and 50\% machine-inferred data (randomly sampled) are used to conduct the multinomial logistic regression and counterfactual analyses
\end{itemize}
There are 5,130 human-annotated tweets after five iterations of the human-guided machine learning process. The remaining tweets of the corpus $C$ are labeled by the machine learning algorithm. According to the previous discussion, 10,945 unique Twitter users are analyzed in the main text. Using the same approach of selecting users with all informative features, (a) 2,939 unique Twitter users inferred only by human-annotated tweets are kept for the robustness verification of the first combination; (b) 6,942 unique Twitter users inferred by both human-annotated and 50\% machine-labeled (randomly sampled) tweets are kept for the robustness verification of the second combination. 

\subsubsection{First combination.}
Descriptive statistics and bi-variate correlations are shown in Table~\ref{tab:char_desc_1st}. Table~\ref{tab:regression_outputs_1st} summarizes the results of the multinomial logistic regression. The Chi-square test shows that the variables significantly predict the opinion on potential COVID-19 vaccines: $\chi^{2}(40, N = 2,939)=399.09, p<.001$, McFadden's pseudo $R^{2}=.07$, which supports our hypotheses.

\begin{sidewaystable*}[htbp]
    \centering
    
    \tabcolsep=0.1cm
    \scriptsize
    \begin{tabular}{l r r l l l l l l l l l l l l l l l l l l l l l l l}
    
    \hline
    \textbf{Variables} & \multicolumn{1}{c}{\textbf{Mean}} & \multicolumn{1}{c}{\textbf{SD}} & \multicolumn{1}{c}{\textbf{1}} &  \multicolumn{1}{c}{\textbf{2}} & \multicolumn{1}{c}{\textbf{3}} &\multicolumn{1}{c}{\textbf{4}} & \multicolumn{1}{c}{\textbf{5}} &\multicolumn{1}{c}{\textbf{6} }&\multicolumn{1}{c}{\textbf{7}} &\multicolumn{1}{c}{\textbf{8}} & \multicolumn{1}{c}{\textbf{9}} &\multicolumn{1}{c}{\textbf{10}} & \multicolumn{1}{c}{\textbf{11}} & \multicolumn{1}{c}{\textbf{12}} &\multicolumn{1}{c}{\textbf{13}}&\multicolumn{1}{c}{\textbf{14}}&\multicolumn{1}{c}{\textbf{15}}&\multicolumn{1}{c}{\textbf{16}}&\multicolumn{1}{c}{\textbf{17}}&\multicolumn{1}{c}{\textbf{18}}&\multicolumn{1}{c}{\textbf{19}}\\
    \hline
    1. Gender (1 = female, 0 = male) & 0.49 & 0.50 \\
    2. Age (years) & 39.02 & 14.81 & -$.10^{**}$\\
    
    3. {\tt Verified} (0 = no, 1 = yes) &0.03 & 0.17 & -$.05^{*}$ & .03 \\
    4. Twitter history (months) & 93.4 & 43.05 & -.02 & $.04^{*}$ & $.09^{**}$\\
    5. \# {\tt Followers} &1.56 & 1.59 & -.00 & $.06^{**}$  & $.40^{**}$ & -$.08^{**}$\\
    6. \# {\tt Friends} & 1.94 & 1.19 & .01 & $.09^{**}$  & $.10^{**}$ & -$.29^{**}$ & $.68^{**}$\\
    7. \# {\tt Listed memberships} & -1.72 & 0.88 & -.03 & $.11^{**}$  & $.54^{**}$ & $.23^{**}$ & $.67^{**}$ & $.30^{**}$\\
    8. \# {\tt Favorites} & 4.38 & 1.71 & $.07^{**}$ & -$.09^{**}$  & -$.05^{**}$ & -$.27^{**}$ & $.30^{**}$ & $.38^{**}$ & .01\\
    9. \# {\tt Statuses} & 4.14 & 1.37& -.01 & -$.10^{**}$  & $.06^{**}$ & -$.08^{**}$ & $.49^{**}$ & $.39^{**}$ & $.25^{**}$ & $.58^{**}$\\
    10. Higher-Income (0 = no, 1 = yes) & 0.01 & 0.07& -.01 & -.03  & $.07^{**}$ & -.01 & $.07^{**}$ & .02 & $.08^{**}$ & .01 & .01\\
    11. Lower-Income (0 = no, 1 = yes) & 0.73 & 0.45 & .01 & -$.50^{**}$  & -.01 & $.08^{**}$ & -$.17^{**}$ & -$.22^{**}$ & -$.10^{**}$ & -$.14^{**}$ & -$.10^{**}$ & -$.12^{**}$\\
    12. Religious (0 = no, 1 = yes) & 0.05 & 0.22 & -.00 & $.10^{**}$  & -.03 & -.01 & $.06^{**}$ & $.10^{**}$ & -.02 & .01 & .02 & -.02 & -$.08^{**}$\\
    13. Having kids (0 = no, 1= yes) &0.12 & 0.32 & $.07^{**}$ & $.09^{**}$  & .02 & -.00 & .03 & .03 & .04 & -.02 & -$.05^{**}$ & -.03 & -$.05^{**}$ & $.13^{**}$\\
    14. Following Trump (0 = no, 1 = yes)& 0.12 & 0.32 & -$.06^{**}$ & $.07^{**}$ & -.02 & -$.18^{**}$ & -$.04^{*}$ & $.05^{*}$ & -$.11^{**}$ & -.01 & -.03 & .02 & -$.05^{**}$ & $.05^{**}$ & $.06^{**}$\\
    15. Following Biden (0 = no, 1 = yes)& 0.15 & 0.36 & $.05^{**}$ & $.08^{**}$ & $.04^{*}$ &  $.10^{**}$ & .02 & $.17^{**}$ & $.05^{**}$ & .03 & -.01 & .02 & -$.06^{**}$ & -.02 & $.07^{**}$ & -.01\\
    16. Rural (0 = no, 1 = yes)& 0.21 & 0.41 & -.01 & $.06^{**}$ & -$.06^{**}$ & -$.07^{**}$ & -$.05^{*}$ & .01 & -$.07^{**}$ & .01 & -.01 & -.00 & -.03 & $.09^{**}$ & $.05^{**}$ & $.05^{**}$ & -.01 \\
    17. Suburban (0 = no, 1 = yes)& 0.14 & 0.35 & -.00 & $.07^{**}$  & -.03 & .02 & -$.06^{**}$ & -.02 & -$.06^{**}$ & -.01 & -.02 & -.03 & -.01 &$.03^{**}$ & .06 & .01 & .02 & -$.21^{**}$ \\
    \makecell[l]{18. Pandemic experience \\ (sentiment)} & 0.01 & 0.78 & $.04^{*}$ & -$.05^{**}$  & $.09^{**}$ & $.04^{*}$ & $.08^{**}$ & .01 & $.15^{**}$ & -$.06^{**}$ & -$.07^{**}$ & .02 & $.04^{*}$ & -$.06^{**}$ & -.02 & -$.06^{**}$ & -.03 & -.02 & -.00\\
    \makecell[l]{19. Non-pandemic experience \\ (sentiment)} & 0.63 & 0.74 & $.09^{**}$ & -$.09^{**}$ & $.07^{**}$ & .04 & $.13^{**}$ & $.08^{**}$ & $.12^{**}$ & $.08^{**}$ & .01 & .02 & $.04^{*}$ & -.01 & .01 & -$.07^{**}$ & -.02 & -$.04^{*}$ & -.01 & $.25^{**}$\\
     \makecell[l]{20. Pandemic severity perception \\ (relative change of \# daily confirmed cases)}& 0.01&0.00& -.01& .03 & -$.06^{**}$ & -.01 & -$.06^{*}$ & -.01 & -$.07^{**}$ & -.02 & -$.04^{*}$ & $.05^{*}$ & .02 & $.06^{**}$ & $.06^{**}$ & .01 &-.02 & $.27^{**}$ & $.15^{**}$ & -.02& .01\\
\hline
    \end{tabular}
    {\raggedright Note. * $p<0.05$. ** $p<0.01$. \par}
    \caption{Descriptive statistics and the bi-variate correlations (Robustness verification of the first combination). The numbers of {\tt followers}, {\tt friends}, {\tt listed memberships}, {\tt favorites}, {\tt statuses} are normalized by the months of Twitter history and log-transformed.}
    \label{tab:char_desc_1st}
\end{sidewaystable*}

\begin{table*}[]
    \centering
    \small
    \setlength{\tabcolsep}{3pt} 
    \renewcommand{\arraystretch}{1.5} 
    \begin{tabular}{r l l l l l l}
    \hline
     & \multicolumn{3}{l}{\textbf{Anti-vaccine}} & \multicolumn{3}{l}{\textbf{Pro-vaccine}}  \\
     \cline{2-7}
     
     Predictor & \textit{B} & \textit{SE} & OR (95\% CI) & \textit{B} & \textit{SE} & OR (95\% CI) \\
\hline
Intercept &-$2.29^{***}$ &0.54 & & -0.15 &0.41 & \\
     Age (years) & $0.01^{**}$ & 0.01 & 1.01 (1.00, 1.02)&  $0.02^{***}$ &0.00  & 1.02 (1.01, 1.03) \\
Twitter history (months) & 0.00 &0.00& 1.00 (1.00, 1.00)& 0.00& 0.00 & 1.00 (1.00, 1.00)\\
           \# {\tt Followers} & $0.32^{***}$  & 0.08 & 1.37 (1.18, 1.60)& 0.12 &0.06& 1.12 (1.00, 1.26)\\
           \# {\tt Friends} & -0.12 & 0.08 & 0.88 (0.75, 1.04) & -0.05 & 0.06 & 0.95 (0.84, 1.08)\\
           \# {\tt Listed memberships} & -$0.56^{***}$ & 0.13 & 0.57 (0.45, 0.73)& 0.03 & 0.09 & 1.03 (0.87, 1.23)\\
           \# {\tt Favorites} & $0.10^{*}$& 0.04 & 1.11 (1.02, 1.21)& $0.26^{***}$& 0.04 & 1.30 (1.21, 1.40)\\
           \# {\tt Statuses} & -0.04 & 0.06 & 0.96 (0.85, 1.07)& -$0.13^{**}$& 0.05& 0.87 (0.80, 0.96)\\
           Pandemic experience (sentiment) & -$0.26^{***}$ & 0.08 & 0.77 (0.66, 0.90)& $0.14^{*}$& 0.06& 1.15 (1.02, 1.30)\\
           Non-pandemic experience (sentiment) & 0.00 & 0.08 &1.00 (0.85, 1.16)& 0.06& 0.07& 1.06 (0.93, 1.20)\\
           \makecell[r]{Pandemic severity perception \\ (relative change of \# daily confirmed cases)} & 6.19 & 15.29 & 486.82 (0.00, 4.99e+15)& -4.12 & 12.43 & 0.02 (0.00, 6.14e+08)\\
           Female & -$0.43^{***}$& 0.12 & 0.65 (0.52, 0.82)& -$0.76^{***}$& 0.10& 0.47 (0.39, 0.56)\\
           {\tt Verified} user & -0.75 &0.61 & 0.47 (0.14, 1.56)& -0.16& 0.32& 0.85 (0.46, 1.59)\\
           Higher-income & -11.86 &325.61 & 0.00 (0.00, 1.03e+272)& 0.71 & 0.68 & 2.03 (0.53, 7.71)\\
           Lower-income & $0.45^{**}$& 0.17& 1.57 (1.14, 2.17)&$0.30^{*}$ & 0.13 &  1.35 (1.06, 1.73)\\
           Religious &$0.84^{**}$ &0.30 & 2.32 (1.28, 4.22)&$0.72^{**}$ &0.27& 2.05 (1.20, 3.52)\\
           Having kids & -0.20& 0.19 & 0.82 (0.57, 1.18)& -0.16& 0.15&0.86 (0.64, 1.14)\\
           Following Trump & $0.71^{***}$& 0.19 &2.04 (1.42, 2.94)& $0.36^{*}$& 0.17&1.43 (1.04, 1.99)\\
           Following Biden & -$1.17^{***}$& 0.19 &0.31 (0.22, 0.45)&-$0.72^{***}$ & 0.13&0.49 (0.38, 0.62)\\
           Rural & $0.42^{**}$& 0.16 & 1.52 (1.12, 2.06)& 0.23 & 0.13 & 1.26 (0.98, 1.62)\\
           Suburban & $0.40^{*}$& 0.17 & 1.49 (1.07, 2.08)& 0.10 & 0.14 &1.11 (0.84, 1.47)\\
          \hline
          Chi-square & \multicolumn{6}{c}{$399.09^{***}$}\\
          \textit{df} & \multicolumn{6}{c}{40}\\
          $-2$ log likelihood & \multicolumn{6}{c}{5,426.51} \\
          McFadden's pseudo $R^{2}$ & \multicolumn{6}{c}{0.07} \\
        Sample size & \multicolumn{6}{c}{2,939}\\          
          \hline
          {\raggedright Note. * $p<0.05$. ** $p<0.01$. *** $p<0.001$.\par}
    \end{tabular}
    \caption{Multinomial logistic regression outputs (Robustness verification of the first combination) for the opinion on potential COVID-19 vaccines against demographics and other variables of interest. The vaccine-hesitant group is selected as the reference category.}
    \label{tab:regression_outputs_1st}
\end{table*}

\begin{figure}[htbp]
    \centering
    \includegraphics[width =\linewidth]{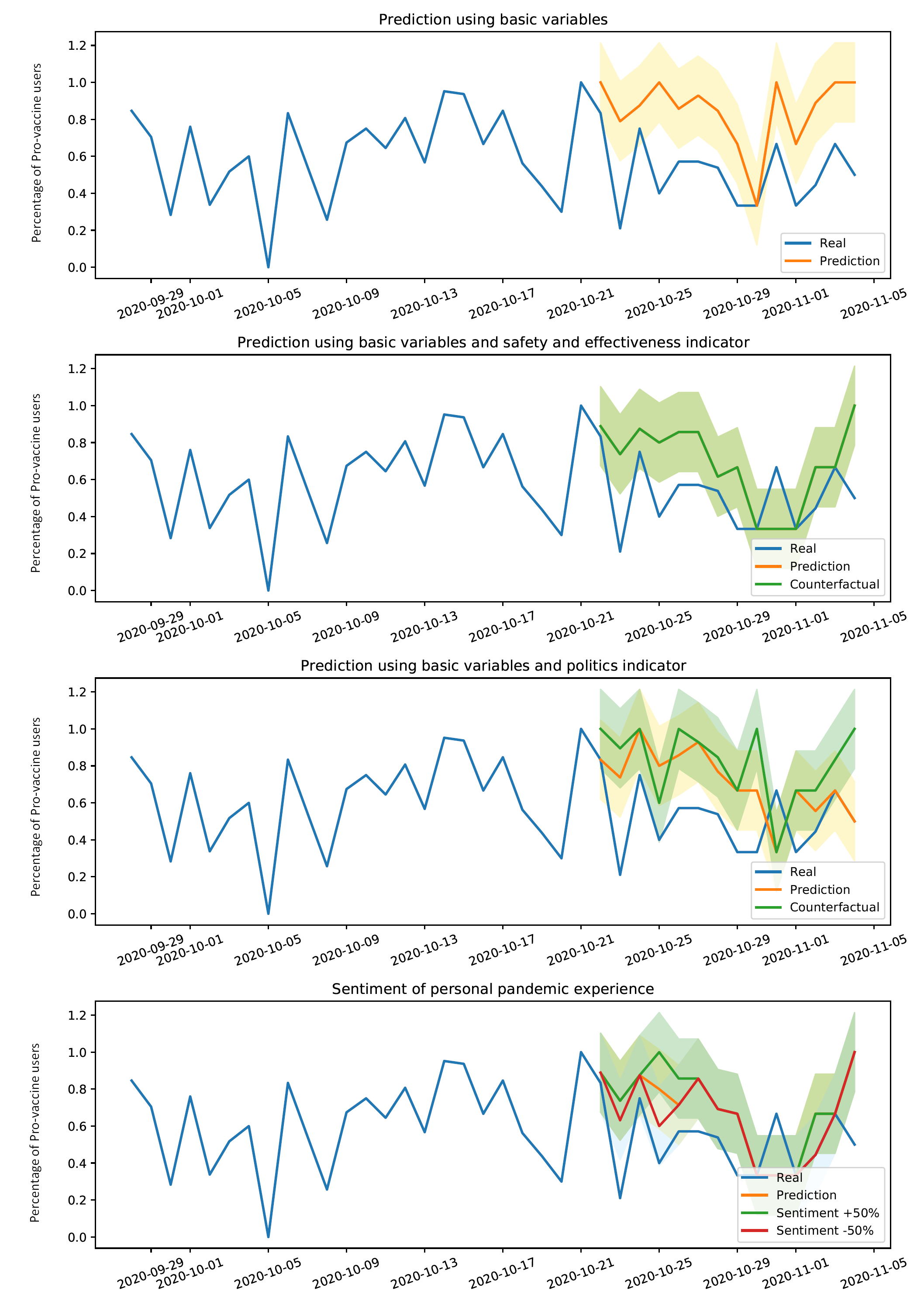}
    \caption{Counterfactual analyses (Robustness verification of the first combination) illustrate the importance of politics, safety and effectiveness factor indicators, and personal pandemic experience.}
    \label{fig:counter_1st}
\end{figure}

Figure~\ref{fig:counter_1st} shows the counterfactual simulation of robustness verification of the first combination. Due to the small sample size, the simulation performance is limited. However, the results do not change fundamentally.

The results of the first robustness verification are basically consistent with the reported results in Table~\ref{tab:regression_outputs} and Figure~\ref{fig:counter}. The findings do not fundamentally change either.

\subsubsection{Second combination.}
Descriptive statistics and bi-variate correlations are shown in Table~\ref{tab:char_desc_2nd}. Table~\ref{tab:regression_outputs_2nd} summarizes the results of the multinomial logistic regression. The Chi-square test shows that the variables significantly predict the opinion on potential COVID-19 vaccines: $\chi^{2}(40, N = 6,942)=807.54, p<.001$, McFadden's pseudo $R^{2}=.06$, which supports our hypotheses.

\begin{sidewaystable*}[htbp]
    \centering
    
    \tabcolsep=0.1cm
    \scriptsize
    \begin{tabular}{l r r l l l l l l l l l l l l l l l l l l l l l l l}
    
    \hline
    \textbf{Variables} & \multicolumn{1}{c}{\textbf{Mean}} & \multicolumn{1}{c}{\textbf{SD}} & \multicolumn{1}{c}{\textbf{1}} &  \multicolumn{1}{c}{\textbf{2}} & \multicolumn{1}{c}{\textbf{3}} &\multicolumn{1}{c}{\textbf{4}} & \multicolumn{1}{c}{\textbf{5}} &\multicolumn{1}{c}{\textbf{6} }&\multicolumn{1}{c}{\textbf{7}} &\multicolumn{1}{c}{\textbf{8}} & \multicolumn{1}{c}{\textbf{9}} &\multicolumn{1}{c}{\textbf{10}} & \multicolumn{1}{c}{\textbf{11}} & \multicolumn{1}{c}{\textbf{12}} &\multicolumn{1}{c}{\textbf{13}}&\multicolumn{1}{c}{\textbf{14}}&\multicolumn{1}{c}{\textbf{15}}&\multicolumn{1}{c}{\textbf{16}}&\multicolumn{1}{c}{\textbf{17}}&\multicolumn{1}{c}{\textbf{18}}&\multicolumn{1}{c}{\textbf{19}}\\
    \hline
    1. Gender (1 = female, 0 = male) & 0.47 & 0.50 \\
    2. Age (years) & 39.62 & 14.64 & -$.08^{**}$\\
    
    3. {\tt Verified} (0 = no, 1 = yes) &0.04 & 0.19 & -.01 & $.03^{*}$ \\
    4. Twitter history (months) & 94.42 & 43.56 & -.02 & $.04^{**}$ & $.09^{**}$\\
    5. \# {\tt Followers} & 1.56 & 1.62 & $.03^{**}$ & -.00  & $.38^{**}$ & -$.07^{**}$\\
    6. \# {\tt Friends} & 1.91 & 1.22& $.04^{**}$ & .01  & $.10^{**}$ & -$.26^{**}$ & $.69^{**}$\\
    7. \# {\tt Listed memberships} & -1.65 & 0.91 & -.01 & $.08^{**}$  & $.51^{**}$ & $.22^{**}$ & $.70^{**}$ & $.33^{**}$\\
    8. \# {\tt Favorites} & 4.17 & 1.88 & $.11^{**}$ & -$.12^{**}$  & -.02 & -$.19^{**}$ & $.36^{**}$ & $.44^{**}$ & $.06^{**}$\\
    9. \# {\tt Statuses} & 4.06 & 1.41& .01 & -$.11^{**}$  & $.07^{**}$ & -$.05^{**}$ & $.51^{**}$ & $.43^{**}$ & $.29^{**}$ & $.57^{**}$\\
    10. Higher-Income (0 = no, 1 = yes) & 0.00 & 0.05 & -.01 & -$.03^{*}$  & $.03^{*}$ & -.02 & $.05^{**}$ & .02 & $.05^{**}$ & .01 & .01\\
    11. Lower-Income (0 = no, 1 = yes) & 0.76 & 0.43 & -.00 & -$.44^{**}$  & -$.04^{**}$ & $.05^{**}$ & -$.15^{**}$ & -$.18^{**}$ & -$.11^{**}$ & -$.14^{**}$ & -$.09^{**}$ & -$.10^{**}$\\
    12. Religious (0 = no, 1 = yes) & 0.04 & 0.19& -.00 & $.07^{**}$  & -$.03^{*}$ & -.02 & $.03^{**}$ & $.06^{**}$ & -.02 & .01 & .01 & .00 & -$.07^{**}$\\
    13. Having kids (0 = no, 1= yes) &0.12 & 0.32 & $.09^{**}$ & $.09^{**}$  & $.03^{**}$ & .02 & $.05^{**}$ & $.05^{**}$ & $.05^{**}$ & .01 & -$.04^{**}$ & -.00 & -$.06^{**}$ & $.09^{**}$\\
    14. Following Trump (0 = no, 1 = yes)& 0.11 & 0.32 & -$.05^{**}$ & $.05^{**}$ & -$.03^{*}$ & -$.18^{**}$ & -$.05^{**}$ & $.04^{**}$ & -$.12^{**}$ & -.02 & -$.05^{**}$ & .00 & -$.04^{**}$ & $.05^{**}$ & $.05^{**}$\\
    15. Following Biden (0 = no, 1 = yes)& 0.17 & 0.37 & $.09^{**}$ & $.07^{**}$ & $.05^{**}$ &  $.10^{**}$ & $.04^{**}$ & $.19^{**}$ & $.05^{**}$ & $.07^{**}$ & .02 & .00 & -$.05^{**}$ & -.02 & $.08^{**}$ & .01\\
    16. Rural (0 = no, 1 = yes)& 0.20 & 0.40 & -.01 & $.08^{**}$ & -$.05^{**}$ & -$.04^{**}$ & -$.06^{**}$ & -.01 & -$.08^{**}$ & -.01 & -$.03^{**}$ & .01 & -$.04^{**}$ & $.06^{**}$ & $.03^{**}$ & $.03^{*}$ & -.02\\
    17. Suburban (0 = no, 1 = yes)& 0.14 & 0.35 & -.00 & $.07^{**}$  & -.02 & -.01 & -$.06^{**}$ & -$.02^{*}$ & -$.05^{**}$ & -$.03^{*}$ & -.02 & -.02 & -.01 &$.03^{**}$ & $.04^{**}$ & $.03^{*}$ & .01 & -$.20^{**}$ \\
    \makecell[l]{18. Pandemic experience \\ (sentiment)} & 0.05 & 0.80 & $.03^{*}$ & -$.05^{**}$  & $.10^{**}$ & $.04^{**}$ & $.10^{**}$ & .02 & $.17^{**}$ & -$.08^{**}$ & -$.07^{**}$ & .02 & $.04^{**}$ & -$.04^{**}$ & -.01 & -$.06^{**}$ & .01 & -.02 & -.01\\
    \makecell[l]{19. Non-pandemic experience \\ (sentiment)} & 0.63 & 0.75 & $.07^{**}$ & -$.07^{**}$ & $.07^{**}$ & $.04^{**}$ & $.14^{**}$ & $.09^{**}$ & $.14^{**}$ & $.06^{**}$ & -.00 & .02 & .02 & -.01 & .02 & -$.06^{**}$ & .01 & -$.03^{*}$ & -$.03^{**}$ & $.27^{**}$\\
     \makecell[l]{20. Pandemic severity perception \\ (relative change of \# daily confirmed cases)}& 0.01&0.00& .00& .01 & -$.05^{**}$ & -$.03^{**}$ & -$.06^{**}$ & -.01 & -$.08^{**}$ & -.01 & -$.04^{**}$ & $.03^{*}$ & -.01 & $.05^{**}$ & $.03^{**}$ & $.03^{*}$ &-$.02^{*}$ & $.26^{**}$ & $.13^{**}$ & -.00& .01\\
\hline
    \end{tabular}
    {\raggedright Note. * $p<0.05$. ** $p<0.01$. \par}
    \caption{Descriptive statistics and the bi-variate correlations (Robustness verification of the second combination). The numbers of {\tt followers}, {\tt friends}, {\tt listed memberships}, {\tt favorites}, {\tt statuses} are normalized by the months of Twitter history and log-transformed.}
    \label{tab:char_desc_2nd}
\end{sidewaystable*}

\begin{table*}[]
    \centering
    \small
    \setlength{\tabcolsep}{3pt} 
    \renewcommand{\arraystretch}{1.5} 
    \begin{tabular}{r l l l l l l}
    \hline
     & \multicolumn{3}{l}{\textbf{Anti-vaccine}} & \multicolumn{3}{l}{\textbf{Pro-vaccine}}  \\
     \cline{2-7}
     
     Predictor & \textit{B} & \textit{SE} & OR (95\% CI) & \textit{B} & \textit{SE} & OR (95\% CI) \\
\hline
Intercept &-$1.97^{***}$ &0.34 & & $0.84^{***}$ &0.26 & \\
     Age (years) & 0.00 & 0.00 & 1.00 (1.00, 1.01)&  $0.01^{***}$ &0.00  & 1.01 (1.01, 1.02) \\
Twitter history (months) & $0.00^{*}$ &0.00& 1.00 (0.997, 1.000)& $0.00^{***}$& 0.00 & 1.00 (1.00, 1.00)\\
           \# {\tt Followers} & $0.31^{***}$  & 0.05 & 1.36 (1.23, 1.50)& $0.08^{*}$ &0.04& 1.08 (1.00, 1.17)\\
           \# {\tt Friends} & -$0.17^{***}$ & 0.05 & 0.84 (0.76, 0.93) & -$0.10^{*}$ & 0.04 & 0.90 (0.83, 0.98)\\
           \# {\tt Listed memberships} & -$0.71^{***}$ & 0.08 & 0.49 (0.42, 0.57)& 0.09 & 0.06 & 1.09 (0.98, 1.22)\\
           \# {\tt Favorites} & 0.04& 0.03& 1.05 (0.99, 1.10)& $0.08^{***}$& 0.02 & 1.08 (1.03, 1.13)\\
           \# {\tt Statuses} & 0.06 & 0.04 & 1.06 (0.99, 1.14)& -$0.08^{**}$& 0.03& 0.93 (0.87, 0.98)\\
           Pandemic experience (sentiment) & -$0.20^{***}$ & 0.05 & 0.82 (0.74, 0.91)& $0.20^{***}$& 0.04& 1.22 (1.13, 1.32)\\
           Non-pandemic experience (sentiment) & -0.06 & 0.05 &0.94 (0.85, 1.04)& $0.10^{*}$& 0.04& 1.10 (1.01, 1.20)\\
           \makecell[r]{Pandemic severity perception \\ (relative change of \# daily confirmed cases)} & -1.02 & 10.02 & 0.36 (0.00, 1.22e+08)& -$20.04^{*}$ & 8.23 & 0.00 (0.00, 0.02)\\
           Female & -$0.28^{***}$& 0.08 & 0.75 (0.65, 0.88)& -$0.51^{***}$& 0.06& 0.60 (0.53, 0.68)\\
           {\tt Verified} user & -0.45&0.34 & 0.64 (0.33, 1.24)& -0.06& 0.18& 0.94 (0.66, 1.34)\\
           Higher-income & -24.13&1.12e+05 & 0.00 (0.00, Inf)& 0.81 & 0.57 & 2.26 (0.73, 6.94)\\
           Lower-income & $0.39^{***}$& 0.11& 1.48 (1.20, 1.82)&$0.45^{***}$ & 0.08 &  1.57 (1.34, 1.84)\\
           Religious &$0.74^{***}$ &0.17 & 2.10 (1.52, 2.91)&$0.46^{*}$ &0.18& 1.58 (1.11, 2.26)\\
           Having kids & -0.11& 0.12 & 0.90 (0.71, 1.14)& -0.13& 0.09&0.88 (0.73, 1.06)\\
           Following Trump & $0.34^{**}$& 0.12 &1.40 (1.11, 1.77)& 0.08& 0.10&1.08 (0.88, 1.32)\\
           Following Biden & -$1.16^{***}$& 0.12 &0.31 (0.25, 0.40)&-$0.36^{***}$ & 0.08&0.70 (0.60, 0.82)\\
           Rural & $0.21^{*}$& 0.10 & 1.24 (1.01, 1.51)& 0.11 & 0.08 & 1.12 (0.95, 1.32)\\
           Suburban & 0.18& 0.11 & 1.20 (0.96, 1.50)& 0.17 & 0.09 &1.18 (0.99, 1.42)\\
          \hline
          Chi-square & \multicolumn{6}{c}{$807.54^{***}$}\\
          \textit{df} & \multicolumn{6}{c}{40}\\
          $-2$ log likelihood & \multicolumn{6}{c}{12,835.77} \\
          McFadden's pseudo $R^{2}$ & \multicolumn{6}{c}{0.06} \\
        Sample size & \multicolumn{6}{c}{6,942}\\          
          \hline
          {\raggedright Note. * $p<0.05$. ** $p<0.01$. *** $p<0.001$.\par}
    \end{tabular}
    \caption{Multinomial logistic regression outputs (Robustness verification of the second combination) for the opinion on potential COVID-19 vaccines against demographics and other variables of interest. The vaccine-hesitant group is selected as the reference category.}
    \label{tab:regression_outputs_2nd}
\end{table*}

\begin{figure}[htbp]
    \centering
    \includegraphics[width = \linewidth]{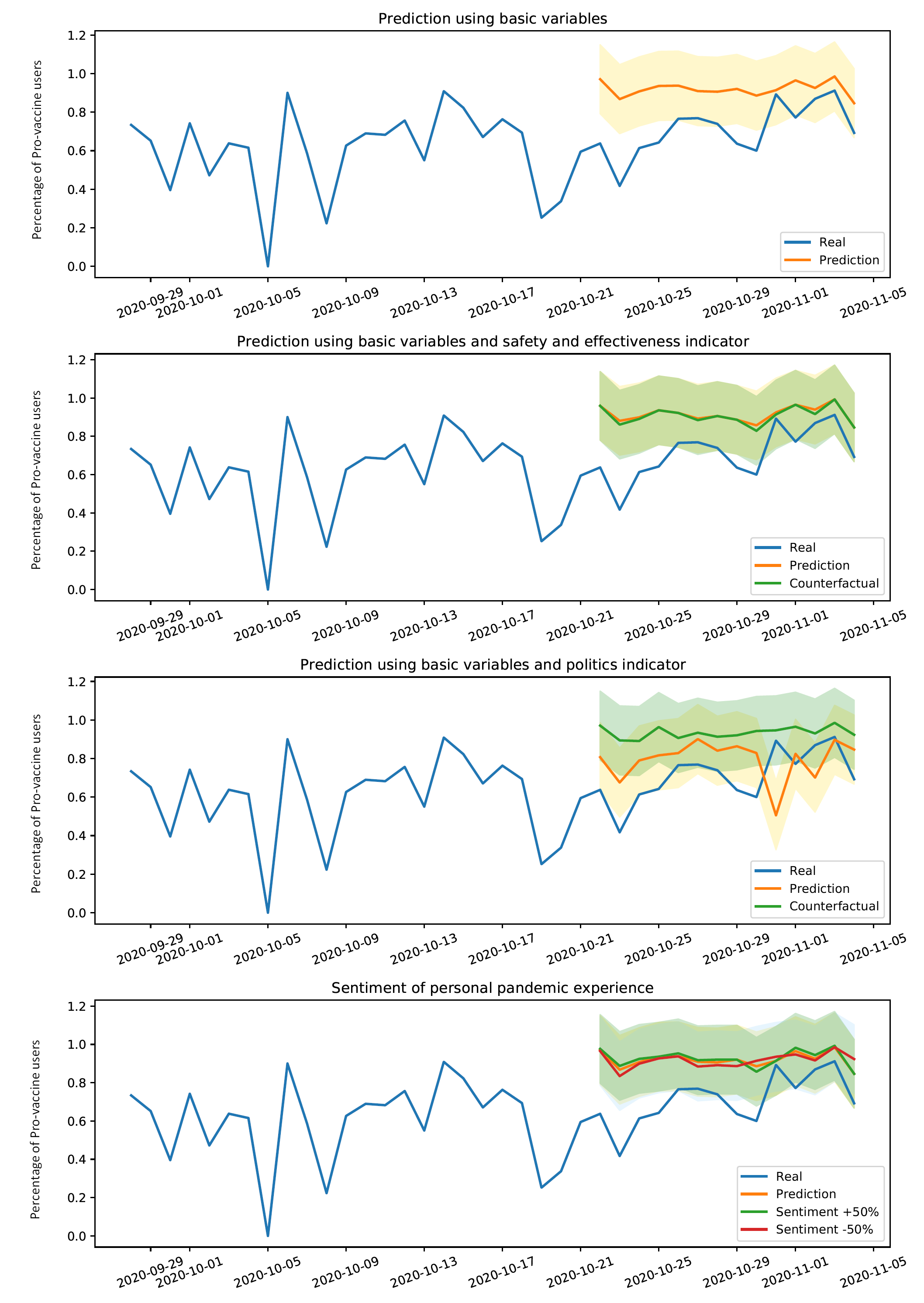}
    \caption{Counterfactual analyses (Robustness verification of the second combination) illustrate the importance of politics, safety and effectiveness factor indicators, and personal pandemic experience.}
    \label{fig:counter_2nd}
\end{figure}

Figure~\ref{fig:counter_2nd} shows the counterfactual simulation of robustness verification of the first combination. The results do not change fundamentally.

The results of the second robustness verification are basically consistent with the reported results in Table~\ref{tab:regression_outputs} and Figure~\ref{fig:counter}. The findings do not fundamentally change either.

\end{document}